\documentclass[aps,prx,twocolumn, longbibliography,notitlepage,superscriptaddress]{revtex4-2}

\usepackage{pdfpages} 
\makeatletter
\AtBeginDocument{\let\LS@rot\@undefined}
\makeatother

\usepackage{subfigure}
\usepackage{textcomp}
\usepackage{graphicx}
\usepackage{bm}
\usepackage{amsmath,amssymb,amsthm}
\usepackage{amsfonts}
\usepackage{dsfont}
\usepackage[colorlinks=true,linkcolor=blue,urlcolor=blue,citecolor=blue]{hyperref}
\usepackage{multirow}
\usepackage{url}
\usepackage[normalem]{ulem}
\usepackage[percent]{overpic}
\usepackage{xcolor}

\newcommand{\beq}{\begin{equation}}
\newcommand{\eeq}{\end{equation}}
\newcommand{\bea}{\begin{eqnarray}}
\newcommand{\eea}{\end{eqnarray}}

\begin{document}
\title{Monitored Fluctuating Hydrodynamics}

\author{Sarang Gopalakrishnan}
\affiliation{Department of Electrical and Computer Engineering, Princeton University, Princeton NJ 08544, USA\\
}

\author{Ewan McCulloch}
\affiliation{Laboratoire de Physique de l'École Normale Supérieure, CNRS,
ENS \& Université PSL; 24 rue Lhomond, 75005 Paris, France}
\affiliation{Department of Electrical and Computer Engineering, Princeton University, Princeton NJ 08544, USA\\
}

\author{Romain Vasseur}
\affiliation{Department of Theoretical Physics, University of Geneva, 24 quai Ernest-Ansermet, 1211 Gen\`eve, Switzerland}

\begin{abstract}

We introduce a hydrodynamic framework for describing monitored classical stochastic processes. We study the conditional ensembles for these monitored processes---i.e., we compute spacetime correlation functions conditioned on a fixed, typical measurement record. In the presence of global symmetries we show that these conditional ensembles can undergo measurement-induced ``sharpening'' phase transitions as a function of the monitoring rate; moreover, even weak monitoring can give rise to novel critical phases, derived entirely from a classical perspective. We give a simple hydrodynamic derivation of the known ``charge-fuzzy phase'' for weakly monitored diffusive many-body quantum systems. We show that although the \emph{unmonitored} symmetric and asymmetric exclusion processes are in different universality classes of transport, the fluctuations in their conditional ensembles flow to the same fixed point with emergent relativistic invariance under monitoring. On the other hand, weakly monitored systems with non-Abelian symmetries enter a novel strongly coupled fixed point with non-trivial dynamical exponent, which we characterize. Our formalism naturally accounts for monitoring general observables, such as currents or density gradients, and allows for a direct calculation of information-theoretic diagnostics of sharpening transitions, including the Shannon entropy of the measurement record.

\end{abstract}

\maketitle

\section{Introduction}
A fundamental problem in statistical physics and information theory is to understand how much an observer can learn about a many-body system from partial measurements of its evolution. This is a problem of inference: given limited access to the system’s degrees of freedom, what features of the underlying probability distribution can be decoded, and how efficiently? Recently this question was explored intensively in the context of monitored quantum dynamics, where local measurements can drive dramatic changes in many-body states. They can destroy correlations, teleport entanglement~\cite{PhysRevLett.70.1895}, or interrupt dynamics through the Zeno effect~\cite{Zeno}. These effects were initially studied in few-qubit systems, but the past decade has seen a shift toward their many-body consequences, motivated by advances in large-scale quantum devices.

A particularly striking discovery in this context was that of measurement-induced phase transitions (MIPTs) in random quantum circuits~\cite{PhysRevX.9.031009, PhysRevB.98.205136,Potter_2022,annurev:/content/journals/10.1146/annurev-conmatphys-031720-030658}. As one increases the rate at which one performs local measurements on a chaotic quantum system, the post-measurement many-body state goes from being highly entangled (and in some ways only weakly affected by measurements) to being weakly entangled~\cite{PhysRevX.10.041020, PhysRevLett.125.030505, PhysRevB.103.104306, PhysRevB.103.174309, PhysRevB.100.134306, PhysRevB.99.224307, PhysRevB.104.104305, PhysRevB.100.064204, PhysRevLett.125.070606, PhysRevB.101.060301, PhysRevB.101.104301, PhysRevB.101.104302,  PhysRevX.11.011030, lavasani_measurement-induced_2021, PhysRevResearch.3.023200,   PhysRevB.103.224210, PhysRevLett.126.060501,PRXQuantum.2.040319, PhysRevB.106.134206, PhysRevLett.126.170503,  PhysRevLett.128.010604, PhysRevB.104.094304, PhysRevResearch.4.023146, PhysRevLett.128.130605,  PhysRevLett.130.220404, PhysRevB.109.125148, PRXQuantum.4.030333,  PhysRevB.110.L060202,  PhysRevB.107.L220204,  PhysRevB.109.014303,   PhysRevB.110.054308, PhysRevLett.130.120402, PhysRevLett.132.240402,  PhysRevB.110.045135, PhysRevB.108.L041103, PhysRevB.110.064323,PhysRevB.107.014308,PhysRevB.107.224303,PhysRevLett.131.060403,PhysRevLett.131.220404,PhysRevB.108.104203,PhysRevB.107.064303, Li_2023,PhysRevB.108.L020306,JianShapourianBauerLudwig2023,
PhysRevX.13.041045,2025arXiv250214034W}.
Although the MIPT was originally explored as a transition in the structure of entanglement, it was subsequently reinterpreted as a transition in what the measuring apparatus learns~\cite{PhysRevB.101.104301,PhysRevLett.129.200602, PRXQuantum.5.020304,agrawal2023observingquantummeasurementcollapse}: in the low-measurement phase, the chaotic dynamics hides quantum information from the measurements~\cite{PhysRevX.10.041020,PhysRevLett.125.030505}, while in the high-measurement phase the measurements efficiently extract information about the initial quantum state of the system. Indeed, a surprising property of these transitions is that they can be detected without explicitly looking at the post-measurement state: they manifest themselves in many-body spacetime correlations of the measurement outcomes, e.g., in the Shannon entropy of the measurement record~\cite{PhysRevLett.128.050602}. 

The interpretation of MIPTs as learnability transitions is particularly transparent when the underlying dynamics conserves a symmetry label~\cite{BAO2021168618,PhysRevX.12.041002,PhysRevB.108.054307,GuoJianFosterLudwigKeldysh2024,MirlinGornyiEtAlKeldysh2024}, such as the global charge of the system~\cite{PhysRevX.12.041002,PhysRevLett.129.120604}. In this setting, a transition occurs in the rate at which an eavesdropper can learn the underlying global symmetry label from local measurements~\cite{PhysRevLett.129.200602,agrawal2023observingquantummeasurementcollapse}: this learning time scales parametrically faster with system size when the measurement rate exceeds a threshold. This transition can be regarded in two ways---as a transition in the learnability of the global labels from the measurement outcomes, or as a transition in the correlation structure of the quantum state that arises conditional on a specific (but typical, i.e., drawn according to their Born probabilities) set of measurement outcomes. For weak measurements, in the slow-learning phase, the charge distribution in this conditional quantum state is ``fuzzy,'' with charge fluctuations over all scales; for strong measurements, in the fast-learning phase, it is ``sharp,'' with the measurement outcomes pinning the charge profile of the conditional state up to local fluctuations confined to a correlation length. 

Recent work has shown that such sharpening transitions do not rely on unitarity, but only on the presence of strong symmetries~\cite{2025arXiv250310308S}. This broadens their scope: they can occur in dissipative quantum processes or even in fully classical stochastic systems, provided that each trajectory respects the conservation law: for example, the electron-phonon coupling has a strong $U(1)$ symmetry because the electrons do not exchange charge with phonons. Since each trajectory of the dynamics preserves the symmetry charge it is meaningful to ask whether, or how fast, the measurements along that trajectory ``learn'' the charge. One can therefore start with a general stochastic process that has a strong symmetry and ask how monitoring modifies inference, without requiring detailed balance or quantum coherence.

From this perspective, our goal is not to engineer a classical analogue of quantum MIPTs, but to recognize that the MIPT framework naturally points to field theories of inference that are equally applicable in classical settings. The classical case is well motivated on its own terms: stochastic processes are ubiquitous in physics, chemistry, and biology, and the task of learning about evolving probability distributions from partial data is a generic challenge—yet the effects of monitoring on such classical processes remain largely unexplored. Moreover, the classical framework avoids the severe practical limitation of post-selection that constrains quantum realizations, offering a simpler and more accessible route to studying learnability transitions.

The present work develops a general hydrodynamic framework for addressing this question, based on a replicated Martin-Siggia-Rose (MSR) formalism~\cite{MSR1,MSR2,MSR3}, and applies it to several paradigm cases of monitored stochastic processes. As a first application, we revisit the problem of charge-sharpening for a single diffusive scalar charge, and provide a simple hydrodynamic derivation of the field theory for the sharpening transition (previously derived by other techniques in Ref.~\cite{PhysRevLett.129.120604}). For this case of the single scalar charge, we demonstrate that one can find new, solvable fixed points in the fuzzy phase, by monitoring the current, the spatial derivative of the charge, etc. We also show that information-theoretic quantities such as the Shannon entropy of the measurement record can be reconstructed from our field theory framework. We then consider a system which is far from equilibrium in the zero-monitoring limit---namely, the asymmetric exclusion process (ASEP)~\cite{SPITZER1970246}---and explore the effects of monitoring. 
We find that although ASEP exhibits KPZ superdiffusive transport in the absence of measurements, upon any nonzero monitoring rate all nonlinear (multi-replica) observables fall into the same universality class of monitored dynamics as systems with diffusive transport. This leaves all physical transport properties unchanged, while strongly modifying the structure of the measurement conditioned ensemble.
(However, some perturbations that were marginal in the diffusive case are irrelevant for ASEP.)
Finally, we turn to the case where monitoring is constrained by a non-Abelian symmetry: as we show, monitoring is a relevant perturbation, which drives the (unmonitored) gaussian fixed point to a new strongly coupled fixed point with a non-trivial dynamical exponent $1<z<2$, which we study numerically and propose a (mean-field) scaling theory for.

The rest of this paper is organized as follows. In Sec.~\ref{setup} we introduce the concepts of strong symmetries, monitored ensembles, and the replica trick for stochastic processes. In Sec.~\ref{secSingleCharge} we revisit the case of a single monitored scalar field, offering a simple hydrodynamic derivation of the field theory for sharpening~\cite{PhysRevLett.129.120604} and extending it to more general cases, e.g., that where the monitored quantity is the current rather than the charge. In Sec.~\ref{burgers}, we turn to the case where the unmonitored dynamics is far from equilibrium, with a current-carrying steady state. Finally, in Sec.~\ref{vector} we turn to monitored vector fields, where the monitored quantity is itself scalar; as we will see this leads to a low-measurement phase described by a new strong-coupling fixed point. 

\section{Setup}\label{setup}

\subsection{Strong symmetries and learning}

We briefly introduce the idea of strong and weak symmetries in the context of classical stochastic processes. These concepts have been discussed extensively for quantum systems~\cite{2012NJPh...14g3007B,PRXQuantum.4.030317,2023arXiv230516399M,PRXQuantum.6.010344,2024PhRvB.110o5150S}, but we emphasize that they are equally well-defined in the classical context. A classical Markov chain is simply the fully dephasing limit of a quantum channel: i.e., one that maps between density matrices that are diagonal in the preferred basis. The definitions of weak and strong symmetries can be read off from this correspondence. In particular, a strong symmetry of the dynamics is defined by the following property, which is easiest to state in terms of the conserved charges associated with the symmetry: a state that initially has a definite value of the conserved charge retains this definite value under strongly symmetric dynamics. By contrast, a weak symmetry just ensures that symmetric probability distributions evolve to other symmetric probability distributions. Thus, a process like electron-phonon coupling has a strong $U(1)$ symmetry associated with charge conservation, while electrons coupled to leads have this symmetry only in the weak sense. 

For any dynamics with a strong symmetry, the problem of charge-sharpening (or learning) is well-defined. One initializes the dynamics in a state of indefinite symmetry charge; one can think of this as the experimentalist's prior uncertainty about the value of the charge. Strongly symmetric dynamics does not introduce any new uncertainty, while measurements yield information and reduce this uncertainty. Eventually the distribution reaches one with definite charge, and after that (because of the strong symmetry) it remains in a state of that definite charge. This reduction in uncertainty due to measurements can be interpreted as learning the global charge. 

\subsection{Trajectory ensembles}

In addition to learning the global symmetry charge, the measurement history also yields information about the local charge distribution in spacetime. To discuss this, it is helpful to introduce a notion of classical trajectory ensembles analogous to the quantum trajectory ensembles that exhibit MIPTs~\cite{2022PhRvB.106u4303P,2022PhRvB.106b4305W,2025arXiv250100547G}. In the setting of this paper, the classical dynamics is generated by a Markov chain, acting on a discrete configuration space, where each configuration is labeled $\vec{x}$ (e.g., if we were considering an exclusion process on $L$ sites, $\vec{x}$ would be a bit-string over $L$ bits). The generalization to continuous time and continuous variables is direct. A classical ``state'' is a probability distribution over $\vec{x}$. Starting from some initial probability distribution and acting for $t$ time steps, the Markov chain generates a joint spacetime distribution $P(\vec{x}_1, \vec{x}_2, \ldots \vec{x}_t) \equiv P(\{ \vec{x}_i \})$. If we monitor the Markov chain at a set of spacetime points and measure a string of outcomes $\mathbf{m}$, the  spacetime history conditioned on these measurement outcomes is $P(\{ \vec{x}_i \} | \mathbf{m})$,  which is given by Bayes' theorem as
\begin{equation} \label{eqBayes}
P(\{ \vec{x}_i \} | \mathbf{m}) = \frac{P(  \mathbf{m} | \{\vec{x}_i \})  P(\{ \vec{x}_i \})}{\sum\nolimits_{\{\vec{y}\}_i} P(  \mathbf{m} | \{\vec{y}_i \})  P(\{ \vec{y}_i \})},
\end{equation}
where the  probability of the measurement outcomes $P(  \mathbf{m} | \{\vec{x}_i \})$ conditional on a given state $\vec{x}$ defines the measurement model. (To monitor the dynamics, one could either weakly condition on every point in spacetime, or completely fix the value of $x_j$ at some set of spacetime points $\{ j \}$. Weak conditioning on site $x_i$ can be formally modeled as taking the value $x_i$, transmitting it through a noisy channel, and conditioning on the output, thus learning a small amount about $x_i$.) For each $\mathbf{m}$, the associated conditional history $P(\{ \vec{x}_i \} | \mathbf{m})$ defines a ``trajectory'' in the sense we will use here. Moreover, each trajectory is associated with a weight that comes from the probability $P(\mathbf{m}) = \sum\nolimits_{\{\vec{x}\}_i} P(  \mathbf{m} | \{\vec{x}_i \})  P(\{ \vec{x}_i \})$ that the corresponding string of measurement outcomes was observed. We are interested in the averages of general functions of $P(\{ \vec{x}_i \} | \mathbf{m})$ over this ensemble, i.e., in quantities of the form  
\begin{equation}\label{ens}
\mathbb{E}_{\mathbf{m}}\left[f(P(\{ \vec{x}_i \} | \mathbf{m}))\right] = \sum\nolimits_{\mathbf{m}} P(\mathbf{m}) f(P(\{ \vec{x}_i \} | \mathbf{m})).
\end{equation}
Notice that for any function linear in $P(\{ \vec{x}_i \} | \mathbf{m})$ (such as an expectation value) the conditioning does not matter: averaging these over trajectories as in Eq.~\eqref{ens} is equivalent to evaluating the expectation value in the state $P(\{\vec{x}_i\})$~\cite{Potter_2022,annurev:/content/journals/10.1146/annurev-conmatphys-031720-030658}. However, even for relatively simple quantities like variances, the trajectory-ensemble average differs from the unconditional average. For example, consider the variance of $x_i$ at some time $t_j$. In the trajectory ensemble, this would be given by the expression $\sum_{\mathbf{m}} P(\mathbf{m}) [\sum\nolimits_{\{\vec{x}\}_i} (x_i(t_j))^2 P(\{ \vec{x}_i \} | \mathbf{m}) - (\sum\nolimits_{\{\vec{x}\}_i} x_i P(\{ \vec{x}_i \} | \mathbf{m}))^2 ]$. The second term is not a linear function of $P(\{ \vec{x}_i \} | \mathbf{m})$, so averages of the form~\eqref{ens} are different from unconditional averages. In general, measures that quantify how much one \emph{learns} from conditioning are nonlinear functions like the variance. To give two examples: charge-sharpening is associated with the vanishing of the charge variance in $P(x_t | \mathbf{m})$, i.e., the state at the final time step. Moreover, charge sharpening is associated with the suppression of local fluctuations in the late-time conditional state $P( \vec{x}_t | \mathbf{m})$: its connected correlation functions, which are covariances, will be short-range in phases that sharpen rapidly and long-range in phases where sharpening is slow.

\subsection{A concrete example: monitored SSEP}
Having discussed monitoring in general stochastic processes, we now describe a concrete example, monitoring the local occupation number in a symmetric simple exclusion process (SSEP). In SSEP, particles undergo excluded hopping in both directions with symmetric rates. We denote each microstate of the system by $|\boldsymbol{s}\rangle$, where $\boldsymbol{s}$ is a string of occupations $s_x \in \{0,1\}$. In the absence of monitoring, the system's state, described by a probability state vector $|P\rangle = \sum_{\boldsymbol{s}} P_{\boldsymbol{s}} |\boldsymbol{s}\rangle$, evolves according to $|P(t)\rangle \equiv e^{-H_{\textrm{SSEP}} t}|P\rangle$, where $H_{\textrm{SSEP}}$ is the generator of the SSEP dynamics~\cite{2015PhyA..418...17M}.

\begin{figure}
    \centering
    \includegraphics[width=0.5\linewidth]{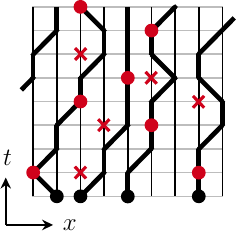}
    \caption{\textbf{Monitored SSEP}: measurement of local occupation at various space-time coordinates constrain the possible state trajectories;
    particle trajectories are forced to avoid unoccupied pins ($\times$) and to pass through occupied pins ($\bullet$).}
    \label{fig:monitoredSSEPcartoon}
\end{figure}

A simple monitored dynamics is to measure occupancy at randomly chosen spacetime coordinates $\{(x,t)\}$. The outcomes, $0$ or $1$, occur with probabilities $1-p$ and $p$ respectively, where $p=\sum_{\boldsymbol{s},s_x=1}\langle \boldsymbol{s} | P(t) \rangle$ is, crucially, linear in $|P(t)\rangle$. Once an outcome $m$ is obtained, the probability vector is projected onto the subspace of consistent configurations, $|P(t)\rangle \to K_m|P(t)\rangle$, where $K_m \equiv \sum_{\boldsymbol{s},s_x = m}|\boldsymbol{s}\rangle \langle \boldsymbol{s}| $. This defines a conditioned ensemble of SSEP trajectories, encoding our knowledge of the system and its history. 

The pinned outcomes act like impurities in disordered systems, constraining possible trajectories (see Fig.~\ref{fig:monitoredSSEPcartoon} for a constrained trajectory in SSEP). A crucial difference is that while quenched disorder is drawn independently of the system’s state, measurement outcomes are sampled from the evolving distribution $|P(t)\rangle$, meaning that one-replica observables such as the average particle current and current statistics are unaffected by monitoring. 

We must distinguish between different types of higher-order correlations here, the higher-moments of transport -- take the full counting statistics of integrated current $J_t$ for example -- is nonlinear in $J_t$, but linear in the trajectory probability. To see this, consider the cumulant generating function of $J_t$ in the unmonitored case, $Z(\lambda) \equiv \mathbb{E}[e^{\lambda J_t}] \equiv \langle \mathds{1}|e^{-tH_{\lambda}}| P(0)\rangle$, where $H_\lambda$ is the tilted SSEP generator with counting field $\lambda$~\cite{2009PhR...478....1T,2007PhRvL..98s5702G,2009JPhA...42g5007G,2007JSP...127...51L,PhysRevLett.131.210402}. Introducing measurements means inserting projectors $K_m$ to obtain a generating function for each measurement history $\boldsymbol{m} = (m_{t_n},\cdots,m_{t_1})$, $Z(\lambda)_{\boldsymbol{m}} = \langle \mathds{1}|e^{-(t-t_{n}) H_{\lambda}}K_{m_{t_n}}\cdots e^{-(t_2-t_1) H_{\lambda}} K_{m_{t_1}} | P(0)\rangle$. Here, $Z(\lambda)_{\boldsymbol{m}}$ is weighted by the likelihood of the outcomes $\boldsymbol{m}$, and we have suppressed the position of measurements in space for brevity. After marginalizing over the measurement outcomes (using $\sum_m K_{m} = 1$), we recover the unmonitored generating function $Z(\lambda) = \sum_{\boldsymbol{m}} Z(\lambda)_{\boldsymbol{m}}$.

Therefore, in order to see the effects of trajectory conditioning, we must use higher-order observables which are truly nonlinear in the state trajectory probabilities, such as the (trajectory averaged) variance in global particle number, $\langle \textrm{Var}(N_{\textrm{tot}})\rangle_{\boldsymbol{m}} = \langle (\langle N^2_{\textrm{tot}}| \hat{P}(t)\rangle - \langle N_{\textrm{tot}}| \hat{P}(t)\rangle^2)\rangle_{\boldsymbol{m}}$ (here $\hat{P}(t)$ is the normalized probability vector for trajectory $\boldsymbol{m}$). As we will see in the following sections, classical measurements (while non-invasive) strongly modify the ensemble of state trajectories, and thus our ability to infer unknown quantities like global charge. In Sec~\ref{burgers} we will return to monitoring exclusion processes (with asymmetric hoppings) in the context of monitoring a Burger's equation.

\subsection{Monitored stochastic evolution and replica trick}

Having introduced the conditional ensembles of interest for a general Markov chain and a stochastic lattice gas, we specialize now to local stochastic processes that can be written as Langevin equations.
Consider a classical stochastic process with degrees of freedom $\rho(\vec{x},t)$, subject to some noise $\xi(\vec{x},t)$. For concreteness, in this paper we will consider systems where $\rho(\vec{x},t)$ is a hydrodynamic field. For example, $\rho(\vec{x},t)$ might be the charge density of a system with a conserved $U(1)$ charge, which in general will obey a fluctuating diffusion equation
\begin{equation} \label{eqDiffusion}
\partial_t \rho  = D \nabla^2 \rho + \nabla \cdot \xi,
\end{equation}
with $\xi$ some Gaussian noise $\langle \xi(x,t) \xi(x',t') \rangle = \sigma^2 \delta(x-x') \delta(t-t')$ whose variance is set by the fluctuation-dissipation theorem. Our formalism can be generalized straightforwardly to any other stochastic classical field theory. 

Now suppose that we weakly monitor some observable $f[\rho(x,t)]$ (this could be the local density, derivatives, products, etc) and condition on the outcome as discussed in the previous section. Denoting $f_{\boldsymbol{m}}(x,t)$
as the likeliest spacetime configuration of $f[\rho(x,t)]$, then, from eq.~\ref{eqBayes}, a general spacetime history $\rho(x,t)$ will be weighted by a factor $P[\rho|f_{\boldsymbol{m}}] \propto P[f_{\boldsymbol{m}}|\rho]= {\rm e}^{- \gamma \int d^d x dt \frac{(f[\rho] - f_{\boldsymbol{m}})^2}{2 }}$ with $\gamma$ the measurement strength. Note also that the Gaussian weighting in the measurement model need not be exact, as corrections to it will generally be irrelevant in the renormalization group sense. For simplicity, we will now assume the simplest case $f[\rho]=\rho$, corresponding to weakly monitoring the field $\rho$ itself, and defer other observables to later sections.

The probability to observe the measurement outcomes $\rho_{\boldsymbol{m}}$ is given by $p[\rho_{\boldsymbol{m}}] = {\mathbb E}_{\rho} {\rm e}^{- \gamma \int d^d x dt \frac{(\rho - \rho_{\boldsymbol{m}})^2}{2 }} \equiv Z[\rho_{\boldsymbol{m}}]/Z  $, where ${\mathbb E}_{\rho}(\dots) = 1/Z \int {\cal D} \rho (...) {\rm e}^{-S[\rho]}$ denotes the average over the solutions of the stochastic process averaged over noise realizations, where the action $S[\rho]$ associated with eq.~\eqref{eqDiffusion} will be written more explicitly below.  
We will denote by 
\begin{align} \label{linearavg}
\langle O(\rho) \rangle_{\rho_{\boldsymbol{m}}} &= \frac{{\mathbb E}_{\rho} O(\rho) {\rm e}^{- \gamma \int d^d x dt \frac{(\rho - \rho_{\boldsymbol{m}})^2}{2 }}}{{\mathbb E}_{\rho} {\rm e}^{- \gamma \int d^d x dt \frac{(\rho - \rho_{\boldsymbol{m}})^2}{2 }}}, \notag \\ 
&= \frac{\int {\cal D} \rho {\rm e}^{-S[\rho]} {\rm e}^{- \gamma \int d^d x dt \frac{(\rho - \rho_{\boldsymbol{m}})^2}{2 }} O(\rho)}{Z[\rho_{\boldsymbol{m}}]} ,
\end{align}
the average over $\rho$ of an observable $O(\rho)$ for fixed measurement outcomes $\rho_{\boldsymbol{m}}$, and denote by $\overline{\langle O(\rho) \rangle_{\rho_{\boldsymbol{m}}} } = \int {\cal D} \rho_{\boldsymbol{m}} p[\rho_{\boldsymbol{m}}] \langle O(\rho) \rangle_{\rho_{\boldsymbol{m}}} $ the average over measurement outcomes. This expression is equivalent to Eq.~\eqref{ens}, and the denominator corresponds to the probability of observing the trajectory corresponding to $\rho_{\boldsymbol{m}}$. As discussed, this average happens to be trivial as the factor $p[\rho_{\boldsymbol{m}}]$ precisely cancels out the denominator in eq.~\eqref{linearavg}.

As we remarked above, non-trivial  measurement-induced phenomena can be probed using non-linear quantities in the trajectories, such as
 \begin{equation}\label{eqnonlinear}
\overline{\langle \rho(x,t) \rangle_{\rho_{\boldsymbol{m}}}  \langle \rho(0,0) \rangle_{\rho_{\boldsymbol{m}}} },
 \end{equation}
 corresponding to a simple class of conditioned classical dynamics. In order to perform the average over the denominator in eq.~\eqref{eqnonlinear}, we use a {\em replica trick}, following recent results for monitored quantum systems~\cite{PhysRevB.99.174205,PhysRevB.100.134203,PhysRevB.101.104301,PhysRevB.101.104302}. We introduce different copies (``replicas'') of the system $\rho_{a}$ with $a=1, \dots ,Q$, each subject to different noise realizations. This allows us to rewrite~\eqref{eqnonlinear} as
 \begin{align}
\lim_{Q \to 1} &\int {\cal D} \rho_{\boldsymbol{m}}  {\mathbb E}_{\rho, \xi} \rho(x,t) {\rm e}^{- \gamma \int d^d x dt \frac{(\rho - \rho_{\boldsymbol{m}})^2}{2 }} \times \notag \\
&{\mathbb E}_{\rho} \rho(0,0) {\rm e}^{- \gamma \int d^d x dt \frac{(\rho - \rho_{\boldsymbol{m}})^2}{2 }} \times p[\rho_{\boldsymbol{m}}]^{Q-2}, \notag \\
&= \lim_{Q \to 1} \langle \rho_1 (x,t) \rho_2(0,0) \rangle_{\rm replicated}.
 \end{align}
The last line denotes an average over the replicated system where the replicas share the same measurement outcomes with weight $\int {\cal D} \rho_{\boldsymbol{m}} {\rm e}^{- \gamma \int d^d x dt \sum_a \frac{(\rho_a - \rho_{\boldsymbol{m}})^2}{2 }}$. We see that monitoring leads to an effective interaction between replicas, with the replica limit $Q \to 1$ ensuring the proper weighting by the trajectory probability $p[\rho_{\boldsymbol{m}}]$. Other non-linear observables characterizing the measurement-induced physics can be obtained through a similar replica procedure, as we will illustrate below.

 \section{Single scalar charge}
\label{secSingleCharge}

In this section we revisit the field theory of charge sharpening for a single scalar charge~\cite{PhysRevLett.129.120604}, but do so using a purely hydrodynamic approach. First, we will introduce this field theory starting from a continuum diffusion equation. This description captures the fuzzy phase at low measurement rate---and admits simple derivations of quantities like correlation functions---but cannot capture the charge-sharp phase at high measurement rate. We discuss simple generalizations of this field theory to measure, not the charge, but other observables like the current. We show how incorporating the discreteness of charges in the underlying stochastic process can be used to restore the sharpening transition and the charge-sharp phase. 
For concreteness, let us first focus on a (linear) diffusive system~\eqref{eqDiffusion}, the generalization to other hydrodynamic systems will be straightforward. 
 
 \subsection{Field theory of the charge-fuzzy phase}

 \subsubsection{Effective replica field theory}
 
 To incorporate the measurements, we will employ the Martin-Siggia-Rose (MSR) formalism~\cite{MSR1,MSR2,MSR3}. Our starting point is the following partition function
 \begin{equation}
 Z = {\mathbb E}_\xi \int  {\cal D} \rho \prod_{\vec{x},t} \delta \left(\partial_t \rho + \nabla. j \right) ,
 \end{equation}
 with the current $j=- D \nabla \rho + \xi $. The physical meaning of this partition function is straightforward: the delta function enforces the diffusion equation and the noise average corresponds to the Gaussian integral ${\mathbb E}_\xi (\dots)= \int {\cal D} \xi {\rm e}^{-\int d^d x dt \xi^2 /(2\sigma^2)} (\dots)$. The variance of the noise correlations is directly related to the diffusion constant through the fluctuation-dissipation theorem $\sigma^2 = 2 \chi D$ with $\chi$ the equilibrium charge susceptibility. In the standard MSR approach, the delta function is implemented using an auxillary field $\varphi$. Integrating out the noise, this leads to an effective field theory for (noisy) diffusion,
 \begin{equation}
 {\cal L} = \frac{\sigma^2}{2} (\nabla \varphi)^2 - i \varphi (\partial_t -D \nabla^2) \rho, 
 \end{equation}
 with $Z = \int {\cal D} \varphi {\cal D} \rho {\rm e}^{- \int d^d x dt {\cal L}}$. This is a Gaussian field theory, and the associated propagators follow from inverting the Gaussian kernel in Fourier space. For example
     \begin{equation} 
 \langle  \rho (\boldsymbol{k},\omega)   \rho(\boldsymbol{k},\omega') \rangle =  \delta(\omega+\omega') \delta \left(\boldsymbol{k} + \boldsymbol{k}' \right)  \frac{(2 \pi)^{d+1}\sigma^2 k^2 }{\omega^2 + k^4 D^2 } .
  \end{equation}
 In the following, we will use the shorthand notation 
     \begin{equation} 
 \langle  \rho   \rho \rangle =  \frac{\sigma^2 k^2 }{\omega^2 + k^4 D^2 } .
  \end{equation}
 for propagators.

We now introduce weak measurements that favor a given space-time trajectory $\rho_{\boldsymbol{m}}(x,t)$
 \begin{equation}
 Z[ \rho_{\boldsymbol{m}}] = {\mathbb E}_\xi \int  {\cal D} \rho \prod_{\vec{x},t} \delta \left(\partial_t \rho + \nabla. j \right) {\rm e}^{- \gamma \int d^d x dt \frac{(\rho - \rho_{\boldsymbol{m}})^2}{2 }} .
 \end{equation}

 The (Born) probability of a given trajectory is given by $p[\rho_{\boldsymbol{m}}] =  Z[ \rho_{\boldsymbol{m}}] / Z $ with $Z=\int {\cal D} \rho_{\boldsymbol{m}} Z[ \rho_{\boldsymbol{m}}]$.

As discussed in the previous section, we  replicate the system $a=1, \dots, Q$, average over measurement outcomes, with physical observables corresponding to $Q\to 1$ in the end 
 \begin{align}
 \overline{Z^Q} &= {\mathbb E}_{\xi^a} \int  {\cal D} \rho_a {\cal D} \rho_{\boldsymbol{m}}  \prod_{a, x,t} \delta \left(\partial_t \rho_a - D \nabla^2 \rho_a - \vec{\nabla} .\vec{ \xi}_a \right)\notag \\
 &\times {\rm e}^{- \gamma \int d^d x dt \sum_a  \frac{(\rho_a - \rho_{\boldsymbol{m}})^2}{2 }} .
 \end{align}
The integral over measurement outcomes is Gaussian and can be performed exactly. We find an effective interaction between replicas
\begin{equation}
{\rm exp} \left( - \gamma \int d^d x dt \sum_{a b}   \frac{\rho_a (\delta_{a,b}-1/Q) \rho_b }{2} \right).
 \end{equation}
Finally, we follow once again the well-known MSR procedure~\cite{MSR1,MSR2,MSR3} to obtain a field theory by writing the delta function as a functional integral over an auxillary response field $\varphi$  and integrating out the noise. We introduce a Lagrangian through $  \overline{Z^Q} = \int  {\cal D} \rho_a  {\cal D} \varphi_a  {\rm exp}\left[- \int d^d x dt {\cal L} \right]$, and find
 \begin{align}
{\cal L} &= \sum_a  \frac{ \sigma^2  }{2 } (\nabla \varphi_a)^2  - i  \varphi_a (\partial_t  - D \nabla^2) \rho_a \notag \\
&+\gamma \sum_{a b} \frac{\rho_a (\delta_{a,b}-1/Q) \rho_b }{2} .
 \end{align}

To deal with the replica structure we introduce replica-symmetric and inter-replica modes. Let $\Pi_{ab} = \delta_{ab} - Q^{-1}$ be the projector onto inter-replica modes, and $\overline{\phi} = \frac{1}{Q} \sum_a \phi_a$ denote the replica-average of the field $\phi$, so that $\phi_a = \overline{\phi} + (\Pi \phi)_a$. The Lagrangian can be decomposed as 
 \begin{align} \label{eqFullLagragian}
 {\cal L}& =  \frac{ \sigma^2  Q }{2 }   ( \nabla \overline{\varphi}) ^2  -i   Q  {\overline \varphi} (\partial_t  - D \nabla^2) \overline{\rho}  \notag \\
&+ \sum_a  \frac{ \sigma^2 }{2 }   ( \nabla \Pi \varphi)_a^2  -i  (\Pi \varphi)_a (\partial_t  - D \nabla^2) (\Pi \rho)_a + \frac{\gamma }{2}  (\Pi \rho)_a^2 .
 \end{align}
In the rest of this paper, we will keep the indices on the $\Pi \rho$ and $\Pi \varphi$ fields implicit.
 The replica-symmetric part of the Lagrangian is just the usual MSR field theory for fluctuating diffusion. This sector is not affected by the measurements. The inter-replica sector is Gaussian, and can be analyzed in a straightforward way. For example, we can integrate out the inter-replica fluctuations of $\rho$ exactly. The measurement term is relevant with RG eigenvalue $y = d+2-d=2$ (the scaling dimension of $\rho$ at the diffusive fixed point is $d/2$, so the perturbation has scaling dimension $d$, to be compared to $d+z = d+2$), corresponding to a crossover scale $\xi_* \sim 1/ \sqrt{\gamma}$, or a crossover time scale $t_* \sim 1/\gamma$.
 To leading order in the gradient expansion, we find the effective action for the inter-replica response field
 \begin{equation}
{\cal L}_{\rm eff} =   \frac{ \sigma^2 }{2 }   ( \nabla \Pi \varphi) ^2 +   \frac{1 }{2\gamma }   ( \partial_t \Pi \varphi) ^2   .
 \end{equation}
 Note that the diffusive Fick term played no role here, only the diffusive noise matters. Of course, note that the diffusion constant is still here through the fluctuation-dissipation theorem $\sigma^2 \sim D$. We  thus find $Q-1$ massless relativistic $z=1$ modes with velocity $v = \sqrt{\sigma^2 \gamma}$. Setting this velocity to one (redefining our space and time units), we have 
 \begin{equation}
{\cal L}_{\rm eff} =   \frac{K }{2 }   ( \partial_\mu \Pi \varphi) ^2 ,
 \end{equation}
 with the stiffness
 \begin{equation}
K \sim \sqrt{\frac{D}{\gamma}}. 
 \end{equation}
 This effective field theory with emergent relatistic invariance describes the so-called ``charge-fuzzy'' phase induced by measurements, that was analyzed in the context of many-body quantum systems in Refs.~\cite{PhysRevX.12.041002,PhysRevLett.129.120604}. Here the same action follows entirely from a classical fluctuating diffusion equation. A key difference with the quantum context is that the field $\varphi$ is (naively) non-compact here, we will come back to this in section~\ref{SecSharpening}. 

\subsubsection{Charge correlators}
  
This field theory framework can be readily used  to compute non-linear observables. For example, using the replica trick, we can write
 \begin{align} \label{eqTimeReplica}
\overline{\langle \rho(\vec{x},t) \rho(0,0) \rangle_{c}   } 
&= \underset{Q \to 1}{\lim} \frac{1}{Q-1} \sum_{a, b}  \langle \rho_a(x,t) \Pi_{ab} \rho_b(0,0) \rangle. 
 \end{align}
Using $\Pi^2 = \Pi$, we have  $\langle \rho_a \Pi_{ab} \rho_b \rangle = \langle (\Pi \rho)_a (\Pi \rho)_a \rangle$. For the (quadratic) Lagrangian~\eqref{eqFullLagragian}, the inter-replica propagators in the $(\Pi \varphi, \Pi \rho)$ basis are given by:
\begin{equation}
G^{-1}=
\begin{pmatrix}
\sigma^{2} k^{2} & \omega - i D k^{2} \\
-\omega - i D k^{2} & \gamma
\end{pmatrix},
\end{equation}
so we have we have
     \begin{equation}\label{eq:free-boson-two-point-func}
\langle (\Pi \rho)_a (\Pi \rho)_b \rangle = \Pi_{ab} \frac{\sigma^2 k^2 }{\omega^2 + k^4 D^2 + \gamma \sigma^2 k^2 } .
  \end{equation}
The replica limit in eq.~\eqref{eqTimeReplica} is thus finite since $\Pi_{aa} = Q-1$ and is given by the propagator of the inter-replica mode $\Pi \rho$.
Equal-time correlations are given by
\begin{equation}
\int d \omega \langle \Pi \rho  \Pi \rho \rangle \sim \frac{| k|}{\sqrt{\gamma}}.
 \end{equation}
This implies that
\begin{equation}
 \overline{\langle \rho(\vec{x},t) \rho(0,t) \rangle_{c}  }\sim \frac{1}{\sqrt{\gamma}}\frac{1}{x^{d+1}}. 
  \end{equation}
The new fixed point's exponents satisfy $d+z = 2 \Delta_{\rho}= 2+2\Delta_\varphi =\Delta_\varphi +\Delta_\rho +z$, with dynamical exponent $z=1$, and the scaling dimensions $\Delta_\rho=(d+1)/2$ and $\Delta_\varphi = (d-1)/2$ of $\Pi \rho$ and $\Pi \varphi$, respectively. 

We can also check that nonlinearities in the original diffusive action are all irrelevant at this new monitored $z=1$ fixed point. For example, the noise variance (conductivity) $\sigma^2 $ depends in general on density $\rho$, so the noise  will generate a term $\rho_a (\nabla \varphi_a)^2$, which will mix replica symmetric and inter-replica modes. One can easily check by power counting that these couplings are all irrelevant and can be ignored.

\subsubsection{Entropy of measurement records}

Our formalism allows for a direct calculation of information theoretic diagnostics. To illustrate this, we consider the Shannon entropy of the measurement record~\cite{PhysRevLett.128.050602}
 \begin{equation}
S = - \int {\cal D} \rho_{\boldsymbol{m}} p[\rho_{\boldsymbol{m}}] \ln p[\rho_{\boldsymbol{m}}],
 \end{equation}
 with $p[\rho_{\boldsymbol{m}}] = Z[\rho_{\boldsymbol{m}}]/Z$. As above, we can use a replica trick to compute this quantity
 \begin{equation}
S = - \underset{{Q \to 1}}{\lim} \frac{d}{dQ} \ln \frac{\int {\cal D} \rho_{\boldsymbol{m}} ( Z[\rho_{\boldsymbol{m}}] )^Q}{Z^Q} ,
 \end{equation}
where we have used $\int {\cal D} \rho_{\boldsymbol{m}} p[\rho_{\boldsymbol{m}}]  = 1$. This quantity is then straightforward to compute using our results above: the replica symmetric mode drops out between the numerator and denominator, and the inter-replica mode is given by $Q-1$ independent free bosons. This gives $S = F_{\rm free \ boson \ CFT}$ where $F_{\rm free \ boson \ CFT}$ is the free energy of the free boson conformal field theory. In particular, in one spatial dimension, the finite-size scaling of this quantity for $ t \gg L$ is governed by the universal form dictated by conformal invariance~\cite{PhysRevLett.56.746,PhysRevLett.56.742,PhysRevLett.128.050602} 
 \begin{equation}
 \frac{S}{t L} = s_\infty  - \frac{\pi v c_{\rm eff}}{6L^2} + \dots,
 \end{equation}
with $v = \sqrt{\sigma^2 \gamma}$ the (non-universal) velocity, and $c_{\rm eff} = 1$ the (universal) effective central charge inherited from the $Q-1$ bosons with central charge $c=1$~\cite{PhysRevB.110.045135}, and $c_{\rm eff} = \underset{{Q \to 1}}{\lim} \frac{d}{dQ} (Q-1) =1$.

 \subsubsection{Simple extensions: Fractons, Levy flights }
 
  The main mechanism for the non-trivial measurement-induced critical phase described above  is the balance between measurements and hydrodynamic noise correlations. In particular, in the case of ordinary diffusion the noise correlations go as $\sim k^2$. Systems with different transport properties can lead to different noise scaling, resulting in a different charge fuzzy phase. For example, we can consider fracton hydrodynamics~\cite{PhysRevLett.125.245303,PhysRevResearch.2.033124,PhysRevX.10.011042} that describe systems where  dipole moment is also conserved, leading to subdiffusive transport with $z=4$. In that case, the noise term is now $\nabla^2 \xi$ corresponding to $\sim k^4$ correlations. Using the formalism above, this immediately predicts a $z=2$ charge-fuzzy phase induced by measurements. This phase is less fuzzy than the diffusive case, with correlations going as $1/x^3$ ($k^2$). Similarly, L\'evy flights~\cite{Levy} with $\sim k^\alpha$ noise correlations will produce tunable $z$ fuzzy phase under monitoring. 
 
 \subsection{Discrete charges and charge sharpening }

\label{SecSharpening}

So far we have taken $\rho$ to be a continuous field,  and our hydrodynamic framework provides a field theory description of the measurement-induced critical charge-fuzzy phase. In monitored many-body quantum systems with a $U(1)$ symmetry, it was shown that at strong measurement rates, the system eventually enters a ``charge-sharp'' phase where charge fluctuations become confined to a finite correlation length~\cite{PhysRevX.12.041002,PhysRevLett.129.120604,GuoJianFosterLudwigKeldysh2024,MirlinGornyiEtAlKeldysh2024}.  To recover this charge-sharpening phase transition in our framework, we reintroduce charge quantization. This can be achieved using the framework of bosonization~\cite{PhysRevLett.129.120604,nahum2025bayesiancriticalpointsclassical}: this leads to compactification effects and topological defects in the effective field theory of the fuzzy phase derived above. Those defects (space-time vortices) can proliferate, leading to a transition into a charge-sharp phase.
As a simple example to see how this compactification emerges, let us take $\rho_{x,t}$ to be a discrete field in space and time, subject to a discrete Langevin equation of the form 
\beq
L_{ij} \rho_j - D_{ij} \xi_j = 0,
\eeq
where the indices are over space and time. All terms in this equation are {\em integer-valued}, including $\rho$ and the noise. 
Following the standard MSR procedure we define the generating function for a given noise realization $Z[\vec{\xi}] = \sum\nolimits_{\vec \rho} \delta(L_{ij} \rho_j - D_{ij} \xi_j)$ as
\beq
Z[\vec{\xi}] = \sum\nolimits_{\vec\rho} \prod\nolimits_i \int_{-\pi}^{\pi} \frac{d \varphi_i}{2 \pi} \exp[ -i \varphi_i (L_{ij} \rho_j - D_{ij} \xi_j)].
\eeq
The next step is the noise average. We will take $P(\xi_i) = (\Delta t/2) \delta(\xi_i = \pm 1) + (1-\Delta t) \delta(\xi_i = 0)$ on every site, and take the time-steps to be small enough, so the noise is close to the identity. The noise term in the MSR action becomes
\beq
\prod\nolimits_i \int_{-\pi}^{\pi} \frac{d \varphi_i}{2\pi} \exp\left[\sum\nolimits_{t, x} \sum_{\alpha=1}^d\Delta t \cos\left(\varphi_{x+e_\alpha}(t) - \varphi_{x}(t)  \right)\right],
\eeq
with $e_\alpha$ the unit vector in spatial direction $e_\alpha$.
Taking a continuum time limit at this stage we find the total action (for a single replica, without measurements)
\begin{align}
S &= \int dt \sum\nolimits_x \left\{ -\sum_{\alpha} \cos  (\varphi_{x+e_\alpha}(t) - \varphi_{x}(t)) \right. \notag \\
& \left. - i \varphi_x(t) \left(\partial_t - D \sum_y \Delta_{xy}\right)  \rho_y(t)\right\},
\end{align}
where $\Delta_{xy}$ is the lattice Laplacian. Taking the continuum limit in space, we recover the action of the continuum case derived above with $-\sum_\alpha\cos  (\varphi_{x+e_\alpha}(t) - \varphi_{x}(t)) \sim \frac{(\nabla \varphi)^2}{2}$ up to irrelevant terms. In particular, the non-linear coupling between replica-average and inter-replica coordinates is irrelevant. The crucial difference with the actions derived directly in the continuum is that the $\varphi$ field is now {\em compact}, and can have topological defects (including space-time vortices in 1+1d). From now on we can treat $\rho$ as continuous again, but with $\varphi$ compact~\cite{GiamarchiBook}.  As argued in Ref.~\cite{PhysRevLett.129.120604}, potential vortex-antivortex pairs in the replica-symmetric sector are strongly confined by  diffusive fluctuations, and can never unbind. Instead, elementary topological defects are in the inter-replica sector, with the minimal example being a composite pair of vortex/antivortex in different replicas. The proliferation of these defects drive a Kotsterlitz-Thouless-like (KT) transition which corresponds to charge-sharpening. A  consequence of this ``vortex-doubling'' is that it halves the critical stiﬀness compared to the ordinary KT transition: $K_c = \frac{1}{\pi}$~\cite{PhysRevLett.129.120604}.
Using standard duality arguments, one can write down a Sine-Gordon-like effective theory for the dual field $\rho \sim \frac{1}{2 \pi} \partial_x \vartheta$~\cite{PhysRevLett.129.120604} (see also~\cite{nahum2025bayesiancriticalpointsclassical}):
 \begin{equation}
{\cal L}_{\rm eff} =   \frac{1 }{8 \pi^2 K }   ( \partial_\mu \Pi \vartheta) ^2 - \lambda \sum_{a \neq b} \cos (\vartheta_a -  \vartheta_b),
 \end{equation}
where ${\rm e}^{i \vartheta_a}$ (resp. ${\rm e}^{-i \vartheta_b}$) inserts a spacetime vortex in replica $a$ (resp.~an antivortex in replica $b$), and $\lambda$ is the inter-replica vortex fugacity. The universal properties of this charge-sharpening transition were studied in Ref.~\cite{PhysRevLett.129.120604}, and we do not discuss them further here. Instead, we will focus on the effective fuzzy phases obtained in the weak monitoring regime in more complicated cases. 

\section{Monitoring gradients and currents}

Our framework is very general, and allows us to monitor different observables. Here we consider monitoring gradients of charge or currents, which reveal less information that measuring the local charge itself, and which are less relevant in the renormalization group sense.

\begin{figure*}[t]
\begin{center}
\includegraphics[width=0.8\textwidth, clip]{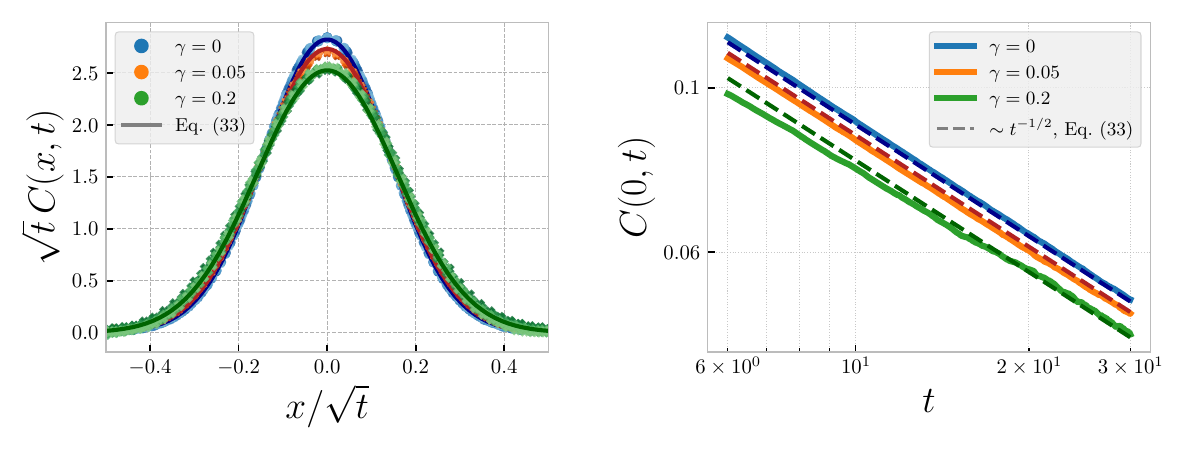}
\caption{{\bf Linear diffusion with gradient monitoring.} (Left) The rescaled dynamical correlator $\sqrt{t}\, C(x,t)$ plotted as a function of $x/\sqrt{t}$ for monitoring strengths $\gamma= 0,\,0.05,\,0.2$ in the monitored noisy linear diffusive system described in Sec.~\ref{gradient-monitoring-numerics}. Consistent with the field-theory prediction Eq.~\eqref{eqn:field-theory-grad-monitoring-corr}, the correlators collapse onto a Gaussian diffusive scaling form, with monitoring increasing the effective diffusion constant and therefore broadening the profile. The solid lines show the theoretical scaling function. (Right) The autocorrelator $C(0,t)$ for the same values of $\gamma$, shown on a log--log scale. In all cases the decay follows the diffusive form $C(0,t)\sim t^{-1/2}$, with a $\gamma$-dependent prefactor $D_{\textrm{eff}}=\sqrt{D^2 + c\,\gamma\,\sigma^2}$
with $c\approx 1.4$, consistent with the renormalized diffusivity in Eq.~\eqref{eqn:field-theory-grad-monitoring-corr}. Numerical data is obtained using the Doob-tilted two-replica stochastic evolution described in Appendix~\ref{Appendix B}.
}
\label{fig:gradient-monitoring}
\end{center}
\end{figure*}

\subsection{Field theory}

Our formalism allows us to consider different types of monitoring. For example, if the observer monitors charge gradient instead of charge density, the inter-replica part of the field theory becomes
 \begin{equation}
{\cal L} =   \frac{ \sigma^2 }{2 }   ( \nabla \Pi \varphi) ^2  -i  (\Pi \varphi) (\partial_t  - D \nabla^2) (\Pi \rho) + \frac{\gamma }{2}  (\Pi \nabla \rho)^2 .
 \end{equation}
 The measurement term is now marginal, so we expect a line of fixed points. The inter-replica propagator becomes
\begin{equation}\label{eqn:field-theory-grad-monitoring-corr}
 \langle \Pi \rho  \Pi \rho \rangle = \frac{\sigma^2 k^2 }{\omega^2 + k^4 (D^2 +\gamma \sigma^2 )},
  \end{equation}
 corresponding to a renormalized diffusion constant
$\tilde{D} = \sqrt{D^2 +\gamma \sigma^2}$.
 In contrast, note that measuring $\partial_t \rho$ leads to an irrelevant perturbation.  

Another natural observable to monitor is the current. In order to deal with this case at the field theory level, we follow the so-called macrosopic fluctuation theory (MFT) perspective~\cite{RevModPhys.87.593}, and trade the noise for the current in the action. For a single unmonitored replica, we have the Lagrangian (focusing on the case of 1d for simplicity)
     \begin{equation} 
{\cal L} = \frac{(j + D \partial_x \rho)^2}{2 \sigma^2} -  i \varphi \left(\partial_t \rho +  \partial_x j \right). 
  \end{equation}
Replicating the action and including a weak monitoring term of the current $\frac{\gamma}{2} \sum_{ab} j_a \left(\delta_{ab} - \frac{1}{Q} \right) j_b$, we obtain a simple quadratic action for the fields. The inter-replica part reads
\begin{equation}
{\cal L} = \frac{(\Pi j + D \partial_x \Pi\rho)^2}{2 \sigma^2} - i \Pi  \varphi \left(\partial_t \Pi \rho +  \partial_x \Pi j \right) + \frac{\gamma}{2} (\Pi j)^2. 
  \end{equation}
 Monitoring currents is a again a marginal perturbation, with current-current correlation 
     \begin{equation} 
 \langle \Pi j  \Pi j \rangle =  \frac{\sigma^2 \omega^2 }{\omega^2 (1 +\gamma \sigma^2) \ + k^4 D^2  } .
  \end{equation}

\subsection{Numerics}\label{gradient-monitoring-numerics}
We now test these predictions numerically in a one dimensional diffusive system subject to weak, noisy monitoring of the lattice gradient of a hydrodynamic charge. In stochastic lattice gases with a bounded on-site charge (e.g.\ $q_x \equiv n_x-\tfrac12 = \pm\tfrac12$ in SSEP or ASEP), the lattice derivative $q_{x+1}-q_x$ is not a purely ``gradient'' observable: the outcomes $0,\pm1$ also reveal $(q_x + q_{x+1})^2$. Thus the lattice derivative has overlap with $q^2$, which is more relevant than $\partial_x q$ at the Gaussian fixed point, and monitoring it constitutes a relevant---rather than marginal---perturbation of diffusive hydrodynamics.

To isolate the effect of genuine gradient monitoring, we instead study a linear fluctuating hydrodynamic model with an \emph{unbounded} charge $\rho_x(t)\in\mathbb{R}$, so that $\rho_{x+1}-\rho_x$ carries no information about $\rho_x^2$. We consider a discretized noisy diffusion equation in which the charge evolves via currents
\begin{equation}
    j_x(t) = -D\,\nabla\rho_x(t) + \xi_x(t),\ \
    \nabla\rho_x(t) \equiv \rho_{x+1}(t) - \rho_x(t),
\end{equation}
with Gaussian white noise $\langle \xi_x(t)\xi_{x'}(t')\rangle
 = 2D\chi^2\,\delta_{x,x'}\delta(t-t')$. We simulate this dynamics in discrete time using a brickwork update: each nearest-neighbour pair $(x,x+1)$ undergoes a charge transfer
\begin{equation}
    \Delta Q_{x,t}
    = -D\,\nabla\rho_{x,t}\,\Delta t
      + \sqrt{\Delta t}\,\xi_{x,t}.
\end{equation}
At each timestep we perform weak measurements of the local gradient, producing noisy readouts $g_{x,t} = \nabla\rho_{x,t} + \eta_{x,t}$, with $\eta_{x,t}$ Gaussian noise of variance $1/\gamma$. Conditioning on a full measurement record $\boldsymbol m$ reweights trajectories according to the Gaussian likelihood $\propto\exp[-\frac{\gamma}{2} \sum_{x,t}\!\left(g_{x,t}-\nabla\rho_{x,t}\right)^{2}]$ of the observed gradients. The dynamical correlation function
\begin{equation}
C(x,t)=
\mathbb{E}_{\boldsymbol m}\!\left[
\langle\rho_{x,t}\rho_{0,0}\rangle_{\boldsymbol m}
- \langle\rho_{x,t}\rangle_{\boldsymbol m}
  \langle\rho_{0,0}\rangle_{\boldsymbol m}
\right],
\end{equation}
is computed by sampling from an equivalent two–replica statistical mechanics model in which trajectories are penalized when the local gradients in the two replicas differ. (This statistical mechanics representation, together with the numerical sampling procedure, is explained in detail in Appendix~\ref{Appendix B}.)

The resulting correlation functions are shown in Fig.~\ref{fig:gradient-monitoring} for $D=1$, $\chi=1$, timestep $\Delta t = 0.01$, and system size $L=200$. For monitoring strengths $\gamma=0,\,0.05,\,0.2$, we find that the autocorrelator $C(0,t)$ decays as $t^{-1/2}$, and the full $C(x,t)$ collapses to a Gaussian diffusive scaling form with a $\gamma$-dependent width, consistent with the renormalized diffusion constant predicted by the effective field theory. A quantitative comparison with Eq.~\eqref{eqn:field-theory-grad-monitoring-corr} yields an effective diffusion constant $D_{\mathrm{eff}} = \sqrt{D^2 + c\,\gamma\,\sigma^2}$ with $c\simeq 1.4$, slightly larger than the Gaussian fixed-point value $c=1$, as expected due to lattice-scale renormalization of the bare couplings (the relation between $D_{\rm eff}$ and the bare couplings is nonuniversal).

\section{Monitored Burgers equation}
\label{burgers}

\begin{figure*}[t]

    \begin{minipage}[t]{0.33\textwidth}
        \begin{overpic}[width=\textwidth,trim=0 1.2cm 0.4cm 1.0cm, clip]{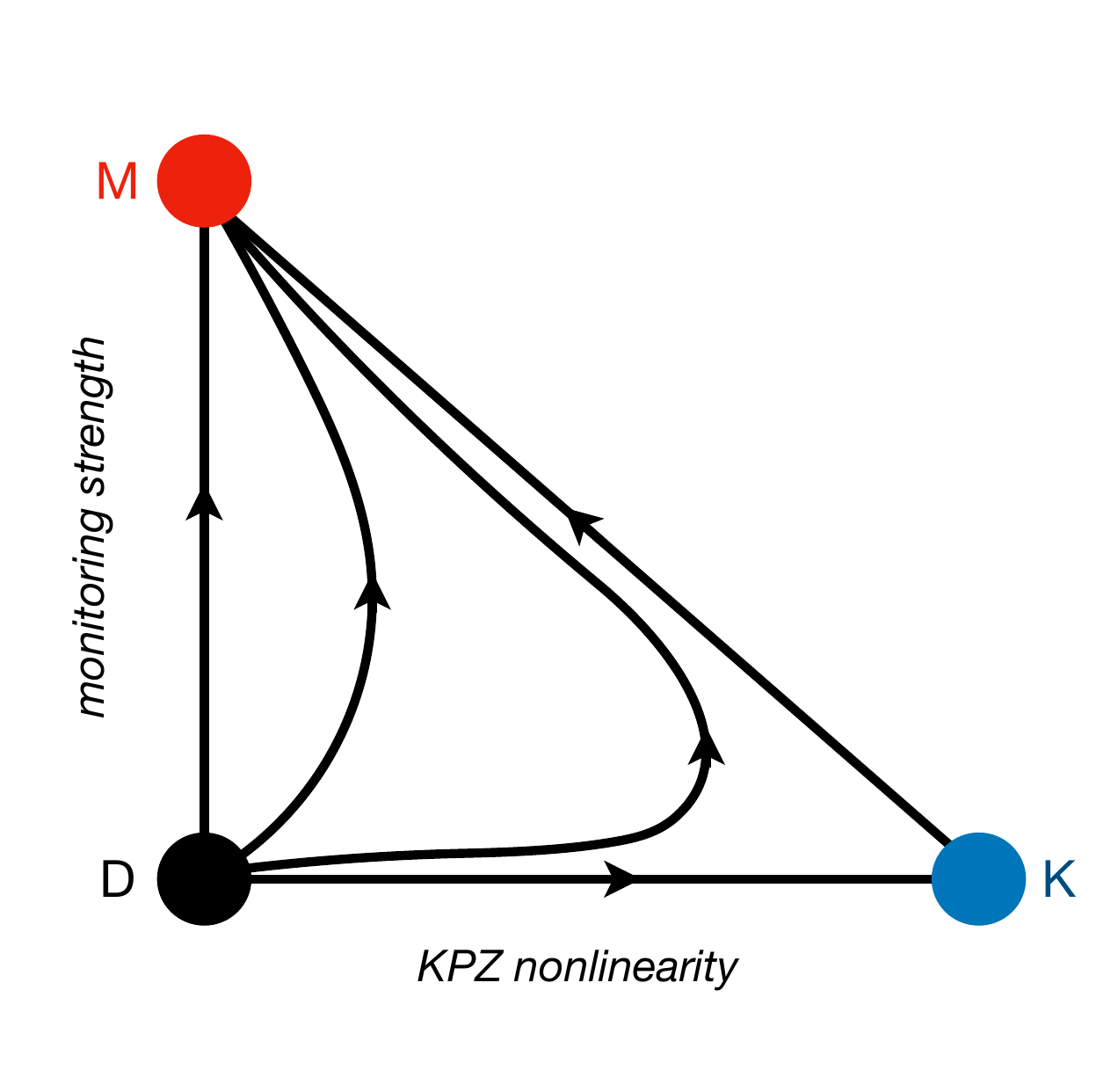}
        \put(-2,80){\large (a)}   
        \end{overpic}
    \end{minipage}
    \hfill
    \begin{minipage}[t]{0.32\textwidth}
        \begin{overpic}[width=\textwidth, trim=1.8cm 1.cm 0.9cm 0.35cm]{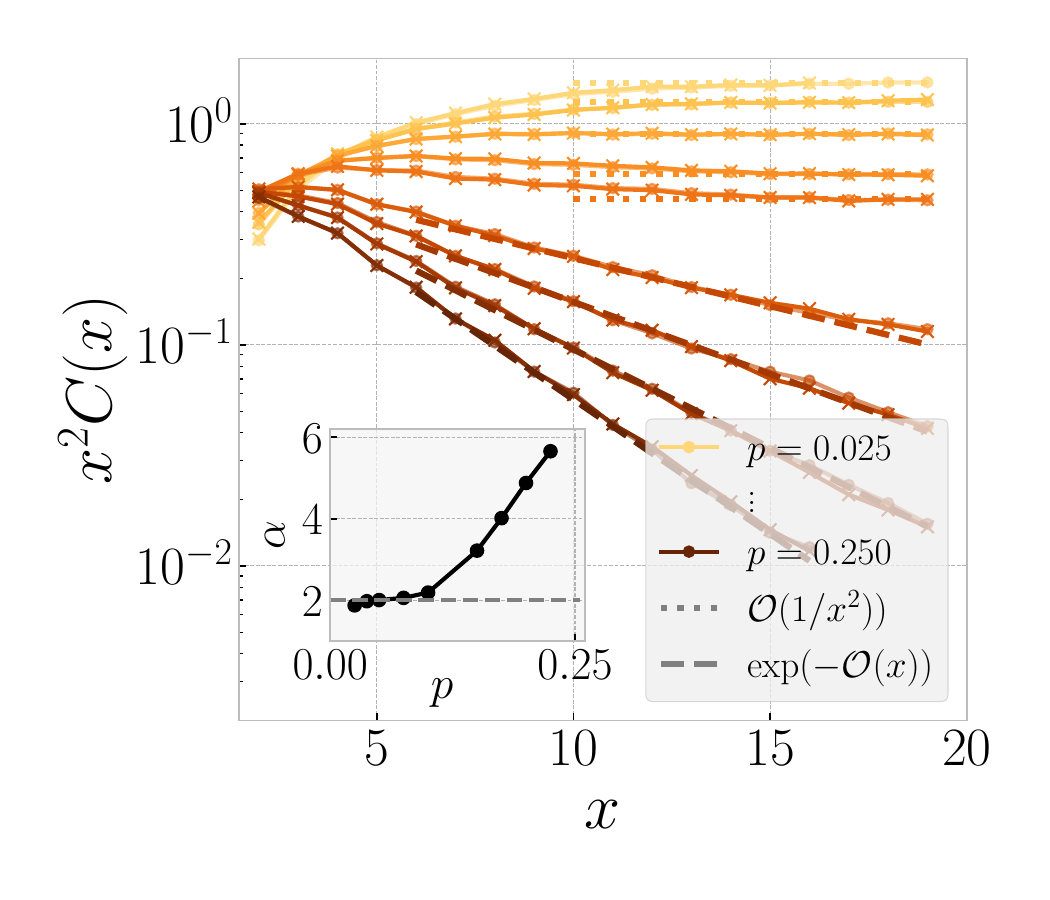}
        \put(-7,80){\large (b)}   
        \end{overpic}
    \end{minipage}
    \hfill
    \begin{minipage}[t]{0.33\textwidth}
        \begin{overpic}[width=\textwidth, trim=0cm 0.cm 0.4cm 0.cm]{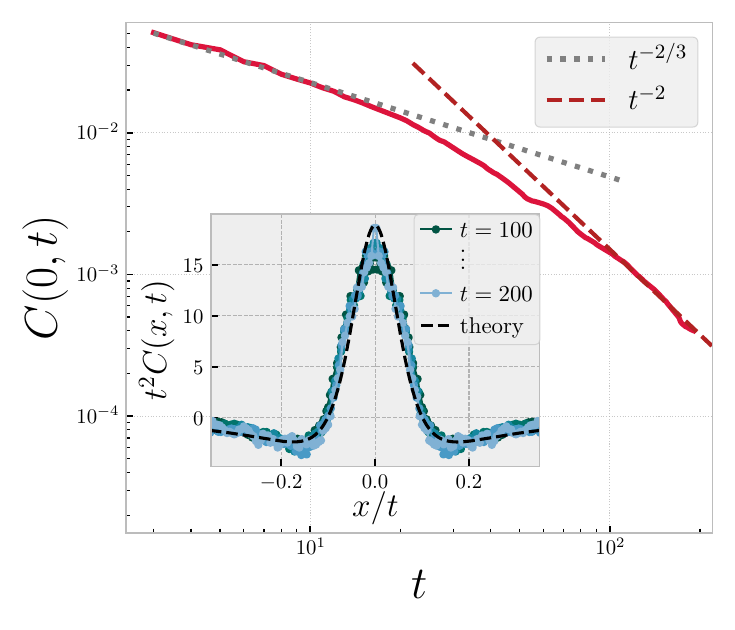}
        \put(1,80){\large (c)}   
        \end{overpic}
    \end{minipage}
    \hfill

\caption{{\bf Burgers' equation with weak monitoring.} (a) The proposed renormalization-group flow shows the unstable fixed points D (diffusion) and K (KPZ), which flow to the stable free-boson fixed point M (monitored) at weak monitoring. For simplicity, we have not shown the strong monitoring phase and the KT-like phase transition here. We are also not depicting the full replica structure -- the diagram depicts only the inter-replica degrees of freedom, the replica symmetric degrees of freedom are un-modified by monitoring. (b) The connected equal-time density-density correlator $C(x)\equiv \mathbb{E}_{\boldsymbol{m}}\left[\langle \rho_x \rho_0\rangle_{\boldsymbol{m}} - \langle \rho_x \rangle_{\boldsymbol{m}} \langle \rho_0\rangle_{\boldsymbol{m}}\right]$ (conditional on a typical measurement , i.e., drawn according to its likelihood) as a function of distance: consistent with the field theory prediction, this correlation function follows a power-law with exponent $\alpha =2 \Delta_\rho$ close to $2$ for weak measurement rate $p$ (main and inset), while for stronger measurements the system enters the sharp phase and these correlations decay exponentially (main). The numerical data are on systems of size $L = 80$ and averaged over $N\approx 1000$ samples (measurement trajectories)~\cite{data}. For each measurement trajectory, the probability state vector is evolved using matrix product state methods with two different maximum bond dimension cutoffs $\chi=200,300$, shown with circle ($\bullet$) and cross ($\times$) markers, to verify convergence. The values of the measurement rate for the various curves are $p \in \{0.025, 0.0375, 0.05, 0.075, 0.1, 0.15, 0.175, 0.2, 0.225\}$ (light to dark). (c) Dynamical two-point function $C(x,t)$ for TASEP under
weak monitoring ($p=0.025$). 
The autocorrelator $C(0,t)$ shows a clear crossover from the KPZ
scaling $C(0,t)\sim t^{-2/3}$ at early times to the predicted
$C(0,t)\sim t^{-2}$ decay at late times (main), consistent with flow to the
stable monitored free-boson fixed point. Furthermore, we observe a scaling collapse of the full dynamical correlation function $C(x,t)$ when plotted as a function of $x/t$ (inset), consistent with the $z=1$ scaling of the free-boson fixed point and showing qualitative agreement with the universal scaling function of Eq.~\eqref{eq:free-boson-two-point-func}.
These simulations sample posteriors for typical measurement histories
using a time-evolving block decimation (TEBD) algorithm with measurement
rate $p=0.025$, system size $L=140$, and bond dimension $\chi=150$.}
\label{fig:monitored-tasep}
\end{figure*}

We now turn to the case where the unmonitored dynamics is far from equilibrium, with a current-carrying steady state, causing anomalous transport due to non-linearities. We consider fluids in one-dimension, described by the stochastic Burgers equation
 \begin{equation}
\partial_t \rho  + \lambda \rho \partial_x \rho = D \partial_x^2 \rho + \partial_x \xi.
 \end{equation}
This equation is directly related to the celebrated Kardar-Parisi-Zhang (KPZ) equation~\cite{KardarParisiZhang1986}, and leads to anomalous sound-waves broadening with dynamical exponent $z=3/2$ (instead of the expected $z=2$ diffusive broadening). Our goal is to study the physics of this system under weak monitoring of charge density. 

\subsection{ Field theory}

Our starting point is the 1d Lagrangian
 \begin{align}
{\cal L} &= \sum_a  \sigma^2 \frac{ (\partial_x \varphi_a)^2 }{2 } - i \sum_a \varphi_a (\partial_t \rho_a  
+ \lambda \rho_a \partial_x \rho_a - D \partial_x^2 \rho_a) \notag \\
&+  \sum_{a b} \gamma \frac{\rho_a (\delta_{a,b}-1/Q) \rho_b }{2} .
 \end{align}
 In the absence of measurements, the KPZ non-linearity $\lambda \varphi \nabla \rho^2 $ is relevant and modifies the dynamical exponent controlling the sound waves' broadening to $z=1+\frac{d}{2}=3/2$, as can be obtained from simple power counting using the fact that the scaling dimensions of $\rho$ and $\lambda$ are fixed to remain $d/2$ to preserve thermal fluctuations. This corresponds to a diffusion constant scaling as $D(k)\sim k^{d/2-1} \sim k^{-1/2}$, or $D(\omega) \sim \omega^{- \frac{1-d/2}{1+d/2}} \sim \omega^{-1/3}$. Note that the noise variance is also scale-dependent $\sigma^2 \sim D \sim k^{-1/2}$, as required by the fluctuation-dissipation theorem. 

We discuss the possible renormalization-group flows in the two-parameter space $(\lambda, \gamma)$. There are three fixed points: the diffusive fixed point ($\lambda = \gamma = 0$), which is perturbatively unstable at small $\lambda$ or small $\gamma$; the KPZ fixed point ($\lambda \to \infty, \gamma = 0$), which is attractive along the $\lambda$ axis; and the free-boson monitored fixed point $M$ ($\gamma = \infty, \lambda = 0$), which is attractive along the $\gamma$ axis. We now consider the stability of the monitored and KPZ fixed points away from the axes. At the free boson fixed point, the KPZ non-linearity has RG eigenvalue $y_\lambda = -1$ and is thus irrelevant. Meanwhile, at the KPZ fixed point, the monitoring term has scaling dimension $\Delta_\gamma=d=1$ and is relevant with RG eigenvalue $y_\gamma=z=3/2$ (at the KPZ fixed point): the corresponding crossover lengthscale is $\xi_* \sim \gamma^{-2/3}$, while the associated timescale is unchanged compared to the simple diffusive case $t_* \sim \gamma^{-1}$. These observations suggest the following simple scenario for the RG flow (see Fig.~\ref{fig:monitored-tasep}(a)): both diffusive and KPZ dynamics flow to the same monitored fixed point, namely the free boson discussed in Sec.~\ref{secSingleCharge}. This shows that different universality classes of transport (diffusion and KPZ) can flow to the same fixed point under weak monitoring. (While we cannot rule out the existence of other fixed points, any consistent modification of this RG flow structure would require introducing at least two new fixed points, and there is no plausible physical mechanism for such fixed points.)

We can also consider monitoring $\partial_x \rho$ at the KPZ fixed point. In that case,  the measurement term now has RG eigenvalue $y_\gamma = -1/2$ and is thus irrelevant (whereas it was marginal in the diffusive case). 

\subsection{Numerics}

We now test this scenario against numerical simulations of a classical stochastic process that is known to be in the KPZ universality class, namely the totally asymmetric exclusion process (TASEP)~\cite{quastel2015one}. TASEP is a stochastic lattice gas with two states on every site (empty and occupied); at each step, a particle can hop to the right with some probability if the site to its right is unoccupied. This stochastic dynamics can be implemented by representing the probability distribution as a matrix product state (MPS) with physical dimension $d=2$ and evolving the MPS using standard techniques~\cite{PhysRevLett.91.147902}. Essentially, we treat it as a quantum state and evolve with imaginary time~\cite{2010PhRvE..82c6702J}. In general, MPS methods are effective for stochastic systems because the stationary states of the corresponding Markov generators (which play the role of ground states) are short-range entangled, and can therefore be represented with modest bond dimension. In our case, the dynamics is punctured by measurements, so the system does not settle into a stationary state. Nevertheless, the imaginary-time evolution still introduces an effective damping that continuously suppresses long-range correlations, allowing much of the many-body state to be truncated with a tiny, controllable error. This ensures that the MPS representation remains faithful throughout the dynamics.

For our simulations, we consider a discrete time TASEP, in which particle hoppings between sites are updated in a brickwork pattern of local updates – i.e., we alternate layers of even-odd and and odd-even gates. At each pair update, a particle attempts a hop right with probability $1/2$, and stays stationary otherwise. Between these layers we randomly draw measurement locations with probability $p$ for each site, stochastically choose an outcome according to the current state probability vector, and project out states inconsistent with this outcome. We take the convention that a single timestep consists of both the even and odd layers of hopping gates, and the intermediary measurements. We initialize a random product state at half-filling (an equal number of empty and occupied sites) as an MPS on $L=80$ sites and run the simulations for times $T \sim L^2$ and over several realizations ($N\approx 1000$) to sample the stationary state and record many shots of non-trivial static correlation functions. 

We work with MPS's of both bond dimension $\chi = 200$ and $\chi = 300$ to verify convergence. For these simulations the truncation error is negligible, so the simulation is essentially exact for the times we study. Imposing periodic boundary conditions on an MPS (needed to maintain a current-carrying steady state in ASEP) is generally a nontrivial task. In the present case we implement periodic boundary conditions by updating the MPS as follows: when a particle at the rightmost site tries to hop right, we make two copies of the MPS: one with the original state, and the other with the state modified so that the rightmost site is empty and the leftmost site is occupied. We then add up these MPS's using standard techniques, and continue the simulation. 

We compute the equal-time particle-number correlations~\eqref{eqTimeReplica} (conditional on a typical measurement record, i.e., drawn according to its likelihood) as a function of distance.
We find clear evidence of a charge-sharpening transition as a function of the measurement rate (see Fig.~\ref{fig:monitored-tasep}(b)). Under weak-monitoring, this correlation function follows a power-law with exponent $\alpha =2 \Delta_\rho$ close to $2$, in agreement with our field theory predictions. For stronger measurements, the system goes into the sharp phase and these correlations decay exponentially. These results indicate that while systems like TASEP and the simple symmetric exclusion process (SSEP) are governed by different transport universality classes (KPZ and diffusion, respectively), they flow to the same infrared fixed point under monitoring.

To further characterize the monitored TASEP fuzzy phase, we compute the dynamical correlation function $C(x,t)$ in the weakly monitored regime ($p=0.025$). For the autocorrelator we observe a clear crossover from a KPZ-dominated regime $C(0,t) \sim t^{-2/3}$ to the asymptotic $t^{-2}$ decay expected at the stable free-boson fixed point (see Fig.~\ref{fig:monitored-tasep}(c)). At late times, the data collapses under the scaling form $C(x,t) = t^{-2} F(x/t)$ with dynamical exponent $z = 1$ (Fig.~\ref{fig:monitored-tasep}(c)), with qualitative agreement with the universal scaling function $F$ predicted in Eq.~\eqref{eq:free-boson-two-point-func}. These results are obtained from posterior sampling of typical measurement trajectories using a TEBD evolution with system size $L = 140$ and bond dimension $\chi = 150$.

  \section{Monitoring non-Abelian charges}

\label{vector}

Finally, we consider a situation with a richer symmetry structure:  imagine that for a symmetry reason, we are only allowed to monitor $\rho^2$ instead of $\rho$. For example, this could be due to a non-Abelian symmetry structure $U(1) \rtimes \mathbb{Z}_2$ where the $\mathbb{Z}_2$ factor acts as particle-hole symmetry $\rho \to - \rho$. We could also imagine a model with vector hydrodynamic charges $\vec{\rho}$, where we are only allowed to measure the scalar quantity $\vec{\rho}^2$. As we will show below, this leads to a new interacting monitored fixed point. 

\begin{figure*}[t]
\begin{center}
\includegraphics[width=\textwidth, trim=3.0cm 0.2cm 0 0, clip]{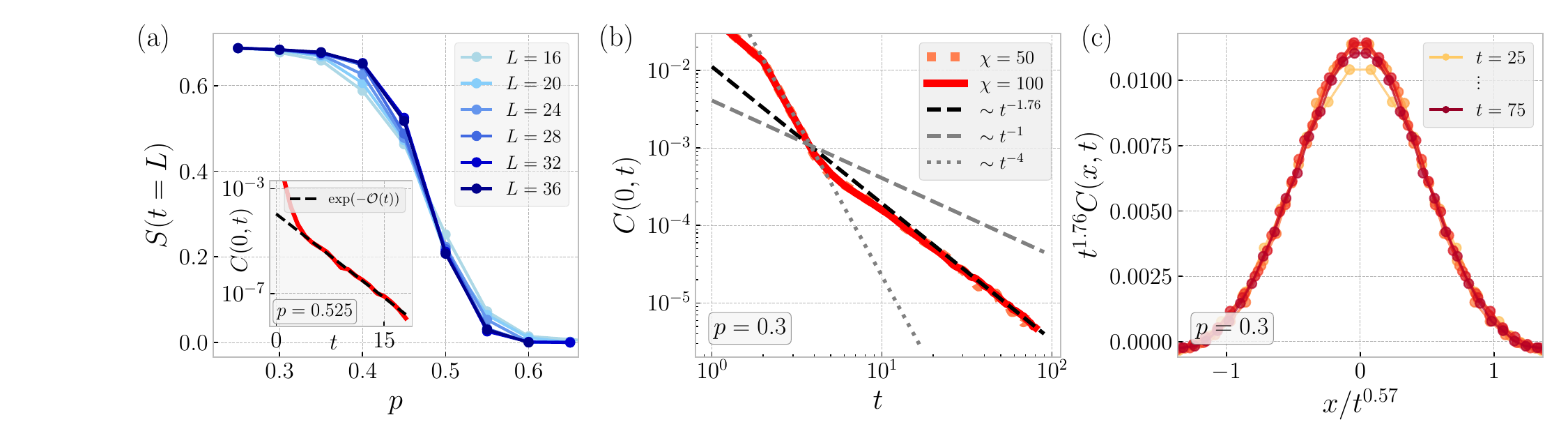}
\caption{{\bf Non-Abelian monitoring.} (a) Matrix product state numerics for the entropy of the posterior charge distribution (see main text) $S(t)$ at time $t = L$ for $L=16,\cdots, 36$ (light to dark) in the three-state stochastic automaton with strong $U(1) \rtimes Z_2$ symmetry. The simulations required bond dimension $\chi = 50$ and are averaged over $N=250$ measurement trajectories. The flow reversal suggests a sharpening transition at $p_{\#}\approx 0.48$, a fuzzy phase ($p < 0.48$) with a sharpening time greater than $L$, i.e., a dynamical exponent $z>1$, and a replica-locked/sharp phase ($p>0.48$). In the replica-locked phase, inter-replica fluctuations are gapped and correlation functions should decay exponentially, as observed for the autocorrelation function of squared local charge density $C(0,t) \equiv \mathbb{E}_{_{\boldsymbol{m}}}\left[\langle \rho^2_0(t) \rho^2_0(0) \rangle_{\boldsymbol{m}} - \langle \rho^2_0(t)\rangle_{\boldsymbol{m}} \langle \rho^2_0(0) \rangle_{\boldsymbol{m}}\right]$ at $p=0.525$ in the sharp phase (inset). (b) The autocorrelator $C(0,t)$ at $p=0.3$ in the fuzzy phase, showing a algebraic decay $C(0,t)\sim t^{-1.76}$. This scaling sits between the abelian $z=1$ fuzzy phase (directly monitoring charge) which has $C(0,t)\sim 1/t^4$ and standard diffusion (the unmonitored $z=2$ phase) which has $C(0,t)\sim 1/t$. These simulations used system size $L=80$, bond dimensions $\chi=50, 100$ to verify convergence, and are averaged over $N=40000$ measurement trajectories. The data is smoothed (with a Gaussian kernel) over a window of size $
\Delta t=2$ to remove persistent oscillations due to the even/odd layer structure of the circuit implementation of the model and the exponent $\alpha=1.76(1)$ is fit over the second half of the evolution time, $t\in [40,80]$. (c) The full dynamical correlation function $C(x,t) \equiv \mathbb{E}_{_{\boldsymbol{m}}}\left[\langle \rho^2_x(t) \rho^2_0(0) \rangle_{\boldsymbol{m}} - \langle \rho^2_x(t)\rangle_{\boldsymbol{m}} \langle \rho^2_0(0) \rangle_{\boldsymbol{m}}\right]$ at $p=0.3$ for times $t=25,\cdots,75$ (light to dark), showing a scaling collapse with non-trivial dynamical exponent $1/z\approx 0.57$ (inset). These simulations used system size $L=80$, bond dimension $\chi=100$, and are averaged over $N=40000$ samples.}
\label{semidirect}
\end{center}
\end{figure*}

\subsection{Field theory and mean-field argument}

The corresponding replica field theory  reads
 \begin{align}
{\cal L} &= \sum_a \frac{ \sigma^2}{2}  (\nabla {\varphi}_a)^2 -i {\varphi}_a (\partial_t  - D \nabla^2)  {\rho}_a \notag \\
&+  \frac{\gamma }{2}\sum_{a b} ({\rho}_a)^2 (\delta_{a,b}-1/Q) ({\rho}_b)^2   .
 \end{align}
 The measurement term now has scaling dimension $2d$, with associated renormalization group (RG) eigenvalue $2-d$. It is relevant in 1d, with upper critical dimension $d=2$. A similar non-linearity will occur for different non-Abelian symmetries. Away from charge neutrality ($\rho=0$) with background density $\rho_*$ , the measurement term will generate terms of the type $\rho_*^2 \rho^2$, where the background is in the replica symmetric sector. At a mean-field level, this again gives a Gaussian action with renormalized measurement strength $\gamma \to \gamma \overline{\rho_*^2}$, and we recover the results of the Abelian case described in the previous section. As we approach charge neutrality $\rho_* \to 0$, we expect a new critical phase with non-trivial dynamical exponent. In the Abelian case, the dynamical exponent can be computed directly from the sharpening time $t_{\sharp}$ -- the time it takes for an observer to learn the global charge of the system. This time scale is set by balancing $\omega^2$ and  $\gamma \sigma^2 k^2$ with $k\sim 1/L$, giving $t_{\sharp} \sim L/\sqrt{\gamma}$. As we approach charge neutrality, the effective monitoring rate of the non-Abelian case $\gamma \to \gamma \overline{\rho_*^2}$ is parametrically reduced. To estimate the corresponding sharpening timescale, we note that replica-symmetric sector is guaranteed to remain diffusive with $z=2$, with Gaussian fluctuations $ \overline{\rho_*^2} \sim 1/L^d$, since it is unaffected by measurements. Using these Gaussian fluctuations of the density, the sharpening time estimated within  mean-field becomes $t_{\sharp} \sim L^{1+d/2}$, or $z=3/2$ in 1d, indicating a non-trivial dynamical exponent. We discuss the stability of this putative strongly interacting fixed point in Appendix.~A and leave a detailed study of this new non-trivial fixed point for future work.

\subsection{Numerics}

To test our analytic theory for the non-Abelian case, we have numerically explored a generalized exclusion process with $U(1) \rtimes Z_2$ symmetry, which is the symmetry resulting from a system with a single conserved charge and a particle hole symmetry after introducing a monitoring protocol which is even in the local charge density. We investigate a stochastic process with states per site $\rho_x = 0,\pm 1$, which undergoes (generalized) SSEP dynamics in the unmonitored case, and then introduce monitoring of the absolute value of local charge (or equivalently, $\rho_x^2$), distinguishing the states $q_x=\pm 1$ from $q_x=0$. We implement a discrete time evolution, which applies layers of even-odd and odd-even gates in which the charge on a pair of sites is left uniformly distributed on the pair. A summary of the transitions $|q_x,q_{x+1}\rangle \to |q'_x,q'_{x+1}\rangle$ is given below,
\begin{align}
|0,0\rangle,\ |\pm1,\mp1\rangle &\to \frac{1}{3}\left(|-1,1\rangle + |0,0\rangle + |1,-1\rangle\right), \nonumber \\
|0,\pm1\rangle,\ |\pm1,0\rangle &\to \frac{1}{2}\left(|0,\pm1\rangle + |\pm1,0\rangle\right),\nonumber \\
|\pm1,\pm1\rangle &\to |\pm1,\pm1\rangle,
\end{align}
Between the even and odd charge hopping layers, $\rho_x^2$ is measured on each site with probability $p$. By simulating the entire state probability vector as a matrix-product state, and using standard techniques to evolve the state~\cite{PhysRevLett.91.147902,2010PhRvE..82c6702J}, we are able to evolve a conditioned probability vector exactly (up to negligible MPS truncation errors). As in the case with monitored TASEP, the outcomes of measurements are drawn from the evolving probability state vector itself (i.e., sampling the true measurement likelihoods) and the conditioning of the probability vector is done by projecting onto the subspace of states consistent with the measurement outcomes. A single timestep $\Delta t = 1$ is chosen to be the following sequence: a layer of even-odd pair gates, a layer of measurements, a layer odd-even of pair gates, and another layer of measurements. Unlike the monitored TASEP, there is no persistent particle/charge current and transport is purely diffusive. We are therefore free to choose open boundary conditions.

To establish the existence of a transition, we explore the entropy of the posterior charge distribution given a typical measurement record and an initial state (prior) that is equally likely to be charge neutral ($\sum_x \rho_x = 0$) or have exactly $+1$ unit of charge ($\sum_x \rho_x = 1$). Specifically, if the initially unknown charge label is $Q$, this is the Shannon entropy of the distribution $P(Q|\mathbf{m})$~\cite{PhysRevLett.129.200602}.
In Fig.~\ref{semidirect}~(a) We plot the remaining posterior entropy as a function of measurement rate for $t \sim L$ and find a crossing at measurement rate $p_{\#} \approx 0.48$: the location of this crossing fixes the sharpening transition, and its existence suggests that the low-measurement phase has $z > 1$ (since larger systems have sharpened less at $t \sim L$).

To further study the putative fuzzy and sharp phases, we investigate the dynamical correlation function of the squared charge density, $C(x,t) \equiv \mathbb{E}_{_{\boldsymbol{m}}}\left[\langle \rho^2_x(t) \rho^2_0(0) \rangle_{\boldsymbol{m}} - \langle \rho^2_x(t)\rangle_{\boldsymbol{m}} \langle \rho^2_0(0) \rangle_{\boldsymbol{m}}\right]$. In the sharp phase (at $p=0.525>p_{\#}$), we find clear exponential decay (see inset of Fig.~\ref{semidirect}~(a)) consistent with gapped inter-replica fluctuations. These simulations used $L=80$, bond dimension $\chi = 100$, and $N=10^6$ samples. 

The approximate (mean-field) power counting argument which gave $z=3/2$ for the fuzzy phase in the previous section also predicts the autocorrelation function scaling $C(0,t)\sim 1/t^2$, which sits between the abelian $z=1$ fuzzy phase which predicts $C(0,t)\sim 1/t^4$, and standard diffusion which has $C(0,t)\sim 1/t$. Numerically, we calculate the dynamical correlation function in the fuzzy phase (at $p=0.3<p_{\#}$) and find an algebraic decay $C(0,t)\sim t^{-\alpha}$ with exponent $\alpha\approx 1.76(1)$ fit over the second half of the evolution time $t\in [40,80]$. We note that the extracted exponent continues to drift slightly with time towards larger values, i.e., away from the abelian sharpening exponent $\alpha = 1$ and towards the mean-field estimate of $\alpha = 2$ for the strongly interacting fixed point for non-abelian monitoring. Furthermore, the full dynamical correlation function shows a scaling collapse with non-trivial dynamical exponent $1/z\approx 0.57$, consistent with the fuzzy phase being described by a fixed point with $1 < z < 2$. These simulations, which use system size $L=80$, bond dimensions $\chi = 50, 100$ (to verify convergence) and are averaged over $N=12500$ measurement trajectories, are shown in Fig.~\ref{semidirect}~(b,c). We defer a detailed exploration of this fuzzy-phase fixed point as well as of the associated sharpening transition to future work.

 \section{Discussion}
\label{SecDiscussion}

In this work we introduced a general field-theory framework for studying information-theoretic phases and phase transitions in monitored stochastic models. These phenomena are properties of the stochastic dynamics conditioned on a particular set of observed measurement outcomes. The main phenomenon we explored was the interplay between the local pinning of the state due to monitoring and the long-range correlations created by slow hydrodynamic modes. As we demonstrated, this new setting of monitored hydrodynamics features new fixed points, even in simple one-dimensional stochastic processes. Many questions remain about the nature of the fixed point we found in the nonabelian case. More generally, the field-theory approach we have introduced here is straightforward to extend to other symmetries, both discrete and continuous. 

The formulation in terms of classical stochastic processes paves the way for systematic explorations of measurement-induced phenomena in dimensions greater than one. Many interesting questions arise specifically in two or more dimensions, and have remained unexplored: for example, the interplay between monitoring and non-reciprocal dynamics in active matter; or the physics of monitored superfluids and liquid crystals. Because many of these systems can be studied in soft-matter settings that do not require quantum coherence, they allow for large-scale experimental studies, which in turn can guide the discovery of new scalable protocols for studying the quantum versions of these phenomena. 

{\it Note.} -- While we were finalizing this work, we became aware of a related work by Nahum and Jacobsen~\cite{nahum2025bayesiancriticalpointsclassical}, which also provides a derivation of the charge-sharpening transition using a replicated MSR framework. The rest of our results largely do not overlap with those of Ref.~\cite{nahum2025bayesiancriticalpointsclassical}.  After our manuscript was posted on arxiv, two additional papers related to classical learnability transitions  and Baysian inference problems appeared~\cite{2025arXiv250408888K,2025arXiv250412385P}.

\section*{Acknowledgments}
\label{Acknowledgments}
The authors thank David Huse, Adam Nahum, Drew Potter, and Yizhi You for helpful discussions and collaborations on related topics. S.G. and E.M. were supported through the Co-design Center for Quantum Advantage
(C2QA) under contract number DE-SC0012704. The simulations presented in this article were performed on computational resources managed and supported by Princeton Research Computing, a consortium of groups including the Princeton Institute for Computational Science and Engineering (PICSciE) and Research Computing at Princeton University.\\

\noindent\textit{All authors contributed equally to this work}.
\\

\section*{Appendix A: Non-abelian monitoring -- dynamical exponent}
\label{Appendix}
In this appendix we characterize the fuzzy phase for non-abelian monitoring more directly at the field theory level. The measurement term couples replica-average and inter-replica modes through terms such as $\sum_b \Pi_{ab} (\Pi {\rho})_b^2 = (\Pi {\rho})_a^2 +  ({ \overline{\rho}})^2 - \overline{({ \rho})^2}$. A crucial point is that if we integrate out inter-replica fluctuations, the remaining theory for the replica-average modes must be trivial with dynamical exponent $\bar{z}=2$ (this sector is unaffected by the measurements). This means that the scaling dimension of  $({ \overline{\rho}})^2 - \overline{({ \rho})^2}$ is guaranteed to remain $d/2$, fixed by Gaussian fluctuations. The measurement part of the action  leads to two types of terms: (1) Inter-replica quartic terms $\sum_a  (\Pi {\rho})_a^4$, and (2) Coupling between inter-replica and replica-symmetric modes such as  $\sim \sum_a  (\Pi {\rho})_a^2 \overline{({ \rho})^2}$ (which we used in the mean-field argument).
 
We can now estimate the dynamical exponent $z$ by requiring that the action (including the interaction) be scale invariant, dropping the (irrelevant) diffusive term $-D \nabla^2$. If we consider the interaction term $\sum_a  (\Pi {\rho})_a^4$ and drop the coupling to the replica-symmetric mode, we have to solve the system of equations $d+z = 4 \Delta_{\rho}= 2+2\Delta_\varphi =\Delta_\varphi +\Delta_\rho +z$, and find $z=(4+d)/3$, so $z=5/3$ in 1d. (Here we have assumed the scaling dimensions satisfy $\Delta_{\rho^{n}}=n\Delta_{\rho}$.) However, at this tentative fixed point, the coupling to the replica symmetric mode $\sim \sum_a  (\Pi {\rho})_a^2 \overline{({ \rho})^2}$ has RG eigenvalue $1/3$, and is thus relevant. If instead we require this coupling to be scale-invariant, the relations give
\begin{equation}\label{non-trivial z}
z = 1 + \frac{d}{2}.
\end{equation}
At this fixed point, the RG eigenvalue of $\sum_a (\Pi {\rho})_a^4$ is $-1/2$ in $d=1$, so it is irrelevant. Thus, we find a self-consistent fixed point at $z=3/2$ in one dimension. We stress, however, that this result should be viewed as a stability check of the mean-field argument under the assumption of canonical scaling for operator dimensions, rather than a controlled derivation. The actual dynamical exponent could differ, but the general structure of the theory strongly suggests a nontrivial value within $1<z<2$.

Physically, one can regard the replica-symmetric mode as diffusively correlated noise, since there is no backaction from the inter-replica modes. In this picture, in one spatial dimension, the effective description resembles a “Luttinger liquid’’ with stiffness $K\sim 1/\sqrt{\overline{\rho}^2}\sim \sqrt{L}$ and velocity $v\sim \sqrt{\overline{\rho}^2}\sim 1/\sqrt{L}$, consistent with $z=3/2$. A more detailed analysis of this nontrivial fixed point is left for future work.

\section*{Appendix B: Gradient monitoring numerics}
\label{Appendix B}

Here we summarize the statistical–mechanics representation underlying the
numerical results of Sec.~\ref{gradient-monitoring-numerics}.  
We consider a discretized noisy diffusion equation for a charge $\rho_x(t)\in \mathbb{R}$, updated in a brickwork scheme, with noisy per-timestep monitoring of the lattice gradients $\nabla\rho_{x,t}$. The measurement at site $x$ and time $t$ yields a noisy readout $g_{x,t} = \nabla\rho_{x,t} + \eta_{x,t}$, with independent Gaussian errors $\eta_{x,t}$ of variance $1/\gamma$. Conditioning on an entire measurement record $\boldsymbol m$ defines a Bayesian posterior over charge trajectories, obtained by reweighting trajectories according to the Gaussian likelihood
\begin{equation}
    W_{\boldsymbol m}(\rho)
    \propto
    e^{
        -\frac{\gamma}{2}
        \sum_{x,t}\!\left(g_{x,t}-\nabla\rho_{x,t}\right)^{2}}.
    \label{eq:posterior-weight-app}
\end{equation}
Since the dynamical two–point function, $C(x,t)=\mathbb{E}_{\boldsymbol m}[
\langle\rho(x,t)\rho(0,0)\rangle_{\boldsymbol m} - \langle\rho(x,t)\rangle_{\boldsymbol m}\langle\rho(0,0)\rangle_{\boldsymbol m}]$, is quadratic in the posterior, it can be rewritten as an \emph{unconditioned} expectation value in an effective two-replica statistical mechanics model. Averaging over the
Gaussian measurement noise yields
\begin{equation}
\mathbb{E}_{\boldsymbol m}
\!\left[
W_{\boldsymbol m}(\rho_1)
W_{\boldsymbol m}(\rho_2)
\right]
\propto
e^{
-\frac{\gamma}{2}\sum_{x,t}
\bigl(\nabla\rho_{1}-\nabla\rho_{2}\bigr)^{2}},
\label{eq:replica-interaction}
\end{equation}
so that trajectories whose gradients disagree are exponentially suppressed. 

It is convenient to regard Eq.~\eqref{eq:replica-interaction} as a Doob tilt of the stochastic dynamics of the two independent replicas.  If $\mathcal L_a$ is the generator of the unmonitored diffusive dynamics for replica $a$ with charge field $\rho_a$, the tilted two-replica generator is
\begin{equation}
    \mathcal{L}^{\textrm{2-rep}}_{\gamma}
    =
    \mathcal L_1 + \mathcal L_2
    - \frac{\gamma}{2}
    \sum_x
      \bigl(\nabla\rho_1 - \nabla\rho_2\bigr)^{2}.
\label{eq:Doob-generator-app}
\end{equation}
Using this representation, the dynamical correlation function can be re-written as
\begin{equation}
C(x,t)
=
\big\langle \rho_1(x,t)\,\rho_1(0,0)\big\rangle_{\textrm{2-rep}}
-
\big\langle \rho_1(x,t)\,\rho_2(0,0)\big\rangle_{\textrm{2-rep}},
\label{eq:C-replica-diff}
\end{equation}
where the expectation is taken with respect to the tilted stochastic evolution generated by Eq.~\eqref{eq:Doob-generator-app}. Thus, computing monitored correlation functions reduces to computing expectation values in the large-deviation ensemble defined by the tilted generator $\mathcal{L}^{\textrm{2-rep}}_{\gamma}$.

The two-replica dynamics is sampled numerically using the population-based Monte–Carlo ``cloning'' algorithm \cite{PhysRevLett.96.120603,2011JSP...145..787G}, which efficiently produces unbiased estimators for observables in exponentially tilted trajectory ensembles. We simulate $n=20000$ samples with $N=1000$ clones each for a system of $L=200$ sites at weak monitoring strength $\gamma=0,0.05,0.02$. The resulting correlation functions are shown in Fig.~\ref{fig:gradient-monitoring}.

\bibliography{MIPT}

@article{PhysRevB.100.134203,
  title = {Entanglement transitions from holographic random tensor networks},
  author = {Vasseur, Romain and Potter, Andrew C. and You, Yi-Zhuang and Ludwig, Andreas W. W.},
  journal = {Phys. Rev. B},
  volume = {100},
  issue = {13},
  pages = {134203},
  numpages = {16},
  year = {2019},
  month = {Oct},
  publisher = {American Physical Society},
  doi = {10.1103/PhysRevB.100.134203},
  url = {https://link.aps.org/doi/10.1103/PhysRevB.100.134203}
}

@article{PhysRevLett.91.147902,
  title = {Efficient Classical Simulation of Slightly Entangled Quantum Computations},
  author = {Vidal, Guifr\'e},
  journal = {Phys. Rev. Lett.},
  volume = {91},
  issue = {14},
  pages = {147902},
  numpages = {4},
  year = {2003},
  month = {Oct},
  publisher = {American Physical Society},
  doi = {10.1103/PhysRevLett.91.147902},
  url = {https://link.aps.org/doi/10.1103/PhysRevLett.91.147902}
}

@article{PhysRevX.12.041002,
  title = {Entanglement and Charge-Sharpening Transitions in U(1) Symmetric Monitored Quantum Circuits},
  author = {Agrawal, Utkarsh and Zabalo, Aidan and Chen, Kun and Wilson, Justin H. and Potter, Andrew C. and Pixley, J. H. and Gopalakrishnan, Sarang and Vasseur, Romain},
  journal = {Phys. Rev. X},
  volume = {12},
  issue = {4},
  pages = {041002},
  numpages = {29},
  year = {2022},
  month = {Oct},
  publisher = {American Physical Society},
  doi = {10.1103/PhysRevX.12.041002},
  url = {https://link.aps.org/doi/10.1103/PhysRevX.12.041002}
}

@article{PhysRevB.101.104302,
  title = {Measurement-induced criticality in random quantum circuits},
  author = {Jian, Chao-Ming and You, Yi-Zhuang and Vasseur, Romain and Ludwig, Andreas W. W.},
  journal = {Phys. Rev. B},
  volume = {101},
  issue = {10},
  pages = {104302},
  numpages = {11},
  year = {2020},
  month = {Mar},
  publisher = {American Physical Society},
  doi = {10.1103/PhysRevB.101.104302},
  url = {https://link.aps.org/doi/10.1103/PhysRevB.101.104302}
}

@article{annurev:/content/journals/10.1146/annurev-conmatphys-031720-030658,
   author = "Fisher, Matthew P.A. and Khemani, Vedika and Nahum, Adam and Vijay, Sagar",
   title = "Random Quantum Circuits", 
   journal= "Annual Review of Condensed Matter Physics",
   year = "2023",
   volume = "14",
   number = "Volume 14, 2023",
   pages = "335-379",
   doi = "https://doi.org/10.1146/annurev-conmatphys-031720-030658",
   url = "https://www.annualreviews.org/content/journals/10.1146/annurev-conmatphys-031720-030658",
   publisher = "Annual Reviews",
   issn = "1947-5462",
   type = "Journal Article",
  }

@article{PhysRevB.99.174205,
  title = {Emergent statistical mechanics of entanglement in random unitary circuits},
  author = {Zhou, Tianci and Nahum, Adam},
  journal = {Phys. Rev. B},
  volume = {99},
  issue = {17},
  pages = {174205},
  numpages = {28},
  year = {2019},
  month = {May},
  publisher = {American Physical Society},
  doi = {10.1103/PhysRevB.99.174205},
  url = {https://link.aps.org/doi/10.1103/PhysRevB.99.174205}
}

@article{Zeno,
    author = {Misra, B. and Sudarshan, E. C. G.},
    title = {The Zeno’s paradox in quantum theory},
    journal = {Journal of Mathematical Physics},
    volume = {18},
    number = {4},
    pages = {756-763},
    year = {1977},
    month = {04},
    issn = {0022-2488},
    doi = {10.1063/1.523304},
    url = {https://doi.org/10.1063/1.523304}
}

@article{PhysRevLett.70.1895,
  title = {Teleporting an unknown quantum state via dual classical and Einstein-Podolsky-Rosen channels},
  author = {Bennett, Charles H. and Brassard, Gilles and Cr\'epeau, Claude and Jozsa, Richard and Peres, Asher and Wootters, William K.},
  journal = {Phys. Rev. Lett.},
  volume = {70},
  issue = {13},
  pages = {1895--1899},
  numpages = {0},
  year = {1993},
  month = {Mar},
  publisher = {American Physical Society},
  doi = {10.1103/PhysRevLett.70.1895},
  url = {https://link.aps.org/doi/10.1103/PhysRevLett.70.1895}
}

@article{SPITZER1970246,
	author = {Frank Spitzer},
	doi = {https://doi.org/10.1016/0001-8708(70)90034-4},
	issn = {0001-8708},
	journal = {Advances in Mathematics},
	number = {2},
	pages = {246-290},
	title = {Interaction of Markov processes},
	url = {https://www.sciencedirect.com/science/article/pii/0001870870900344},
	volume = {5},
	year = {1970}}

@ARTICLE{2025arXiv250310308S,
       author = {{Singh}, Hansveer and {Vasseur}, Romain and {Potter}, Andrew C. and {Gopalakrishnan}, Sarang},
        title = "{Mixed-state learnability transitions in monitored noisy quantum dynamics}",
      journal = {arXiv e-prints},
         year = 2025,
        month = mar,
          eid = {arXiv:2503.10308},
        pages = {arXiv:2503.10308},
          doi = {10.48550/arXiv.2503.10308},
archivePrefix = {arXiv},
       eprint = {2503.10308},
 primaryClass = {quant-ph},
       adsurl = {https://ui.adsabs.harvard.edu/abs/2025arXiv250310308S}
}

@article{PhysRevB.101.104301,
  title = {Theory of the phase transition in random unitary circuits with measurements},
  author = {Bao, Yimu and Choi, Soonwon and Altman, Ehud},
  journal = {Phys. Rev. B},
  volume = {101},
  issue = {10},
  pages = {104301},
  numpages = {26},
  year = {2020},
  month = {Mar},
  publisher = {American Physical Society},
  doi = {10.1103/PhysRevB.101.104301},
  url = {https://link.aps.org/doi/10.1103/PhysRevB.101.104301}
}

@article{PhysRevB.100.134306,
  title = {Measurement-driven entanglement transition in hybrid quantum circuits},
  author = {Li, Yaodong and Chen, Xiao and Fisher, Matthew P. A.},
  journal = {Phys. Rev. B},
  volume = {100},
  issue = {13},
  pages = {134306},
  numpages = {26},
  year = {2019},
  month = {Oct},
  publisher = {American Physical Society},
  doi = {10.1103/PhysRevB.100.134306},
  url = {https://link.aps.org/doi/10.1103/PhysRevB.100.134306}
}

@article{PhysRevB.101.060301,
  title = {Critical properties of the measurement-induced transition in random quantum circuits},
  author = {Zabalo, Aidan and Gullans, Michael J. and Wilson, Justin H. and Gopalakrishnan, Sarang and Huse, David A. and Pixley, J. H.},
  journal = {Phys. Rev. B},
  volume = {101},
  issue = {6},
  pages = {060301},
  numpages = {5},
  year = {2020},
  month = {Feb},
  publisher = {American Physical Society},
  doi = {10.1103/PhysRevB.101.060301},
  url = {https://link.aps.org/doi/10.1103/PhysRevB.101.060301}
}

@article{PhysRevX.10.041020,
  title = {Dynamical Purification Phase Transition Induced by Quantum Measurements},
  author = {Gullans, Michael J. and Huse, David A.},
  journal = {Phys. Rev. X},
  volume = {10},
  issue = {4},
  pages = {041020},
  numpages = {28},
  year = {2020},
  month = {Oct},
  publisher = {American Physical Society},
  doi = {10.1103/PhysRevX.10.041020},
  url = {https://link.aps.org/doi/10.1103/PhysRevX.10.041020}
}

@article{PhysRevB.109.014303,
  title = {Boundary transfer matrix spectrum of measurement-induced transitions},
  author = {Kumar, Abhishek and Aziz, Kemal and Chakraborty, Ahana and Ludwig, Andreas W. W. and Gopalakrishnan, Sarang and Pixley, J. H. and Vasseur, Romain},
  journal = {Phys. Rev. B},
  volume = {109},
  issue = {1},
  pages = {014303},
  numpages = {16},
  year = {2024},
  month = {Jan},
  publisher = {American Physical Society},
  doi = {10.1103/PhysRevB.109.014303},
  url = {https://link.aps.org/doi/10.1103/PhysRevB.109.014303}
}

@article{PhysRevLett.125.030505,
  title = {Quantum Error Correction in Scrambling Dynamics and Measurement-Induced Phase Transition},
  author = {Choi, Soonwon and Bao, Yimu and Qi, Xiao-Liang and Altman, Ehud},
  journal = {Phys. Rev. Lett.},
  volume = {125},
  issue = {3},
  pages = {030505},
  numpages = {6},
  year = {2020},
  month = {Jul},
  publisher = {American Physical Society},
  doi = {10.1103/PhysRevLett.125.030505},
  url = {https://link.aps.org/doi/10.1103/PhysRevLett.125.030505}
}

@article{PhysRevLett.128.050602,
  title = {Operator Scaling Dimensions and Multifractality at Measurement-Induced Transitions},
  author = {Zabalo, A. and Gullans, M. J. and Wilson, J. H. and Vasseur, R. and Ludwig, A. W. W. and Gopalakrishnan, S. and Huse, David A. and Pixley, J. H.},
  journal = {Phys. Rev. Lett.},
  volume = {128},
  issue = {5},
  pages = {050602},
  numpages = {6},
  year = {2022},
  month = {Feb},
  publisher = {American Physical Society},
  doi = {10.1103/PhysRevLett.128.050602},
  url = {https://link.aps.org/doi/10.1103/PhysRevLett.128.050602}
}

@article{PhysRevB.103.104306,
  title = {Statistical mechanics of quantum error correcting codes},
  author = {Li, Yaodong and Fisher, Matthew P. A.},
  journal = {Phys. Rev. B},
  volume = {103},
  issue = {10},
  pages = {104306},
  numpages = {19},
  year = {2021},
  month = {Mar},
  publisher = {American Physical Society},
  doi = {10.1103/PhysRevB.103.104306},
  url = {https://link.aps.org/doi/10.1103/PhysRevB.103.104306}
}

@article{PhysRevLett.125.070606,
  title = {Scalable Probes of Measurement-Induced Criticality},
  author = {Gullans, Michael J. and Huse, David A.},
  journal = {Phys. Rev. Lett.},
  volume = {125},
  issue = {7},
  pages = {070606},
  numpages = {6},
  year = {2020},
  month = {Aug},
  publisher = {American Physical Society},
  doi = {10.1103/PhysRevLett.125.070606},
  url = {https://link.aps.org/doi/10.1103/PhysRevLett.125.070606}
}

@article{PhysRevLett.129.200602,
  title = {Transitions in the Learnability of Global Charges from Local Measurements},
  author = {Barratt, Fergus and Agrawal, Utkarsh and Potter, Andrew C. and Gopalakrishnan, Sarang and Vasseur, Romain},
  journal = {Phys. Rev. Lett.},
  volume = {129},
  issue = {20},
  pages = {200602},
  numpages = {7},
  year = {2022},
  month = {Nov},
  publisher = {American Physical Society},
  doi = {10.1103/PhysRevLett.129.200602},
  url = {https://link.aps.org/doi/10.1103/PhysRevLett.129.200602}
}

@article{quastel2015one,
  title={The one-dimensional KPZ equation and its universality class},
  author={Quastel, Jeremy and Spohn, Herbert},
  journal={Journal of Statistical Physics},
  volume={160},
  pages={965--984},
  year={2015},
  publisher={Springer}
}

@article{PhysRevLett.129.120604,
  title = {Field Theory of Charge Sharpening in Symmetric Monitored Quantum Circuits},
  author = {Barratt, Fergus and Agrawal, Utkarsh and Gopalakrishnan, Sarang and Huse, David A. and Vasseur, Romain and Potter, Andrew C.},
  journal = {Phys. Rev. Lett.},
  volume = {129},
  issue = {12},
  pages = {120604},
  numpages = {7},
  year = {2022},
  month = {Sep},
  publisher = {American Physical Society},
  doi = {10.1103/PhysRevLett.129.120604},
  url = {https://link.aps.org/doi/10.1103/PhysRevLett.129.120604}
}

@article{KardarParisiZhang1986,
  title = {Dynamic Scaling of Growing Interfaces},
  author = {Kardar, Mehran and Parisi, Giorgio and Zhang, Yi-Cheng},
  journal = {Physical Review Letters},
  volume = {56},
  number = {9},
  pages = {889--892},
  year = {1986},
  month = {March},
  publisher = {American Physical Society},
  doi = {10.1103/PhysRevLett.56.889},
  url = {https://journals.aps.org/prl/abstract/10.1103/PhysRevLett.56.889}
}

@article{PhysRevX.9.031009,
  title = {Measurement-Induced Phase Transitions in the Dynamics of Entanglement},
  author = {Skinner, Brian and Ruhman, Jonathan and Nahum, Adam},
  journal = {Phys. Rev. X},
  volume = {9},
  issue = {3},
  pages = {031009},
  numpages = {21},
  year = {2019},
  month = {Jul},
  publisher = {American Physical Society},
  doi = {10.1103/PhysRevX.9.031009},
  url = {https://link.aps.org/doi/10.1103/PhysRevX.9.031009}
}

@ARTICLE{2025arXiv250214034W,
       author = {{Wang}, Qingyuan and {Vasseur}, Romain and {Trebst}, Simon and {Ludwig}, Andreas W.~W. and {Zhu}, Guo-Yi},
        title = "{Decoherence-induced self-dual criticality in topological states of matter}",
      journal = {arXiv e-prints},
         year = 2025,
        month = feb,
          eid = {arXiv:2502.14034},
        pages = {arXiv:2502.14034},
          doi = {10.48550/arXiv.2502.14034},
archivePrefix = {arXiv},
       eprint = {2502.14034},
 primaryClass = {quant-ph},
       adsurl = {https://ui.adsabs.harvard.edu/abs/2025arXiv250214034W}
}

@article{PhysRevB.98.205136,
  title = {Quantum Zeno effect and the many-body entanglement transition},
  author = {Li, Yaodong and Chen, Xiao and Fisher, Matthew P. A.},
  journal = {Phys. Rev. B},
  volume = {98},
  issue = {20},
  pages = {205136},
  numpages = {9},
  year = {2018},
  month = {Nov},
  publisher = {American Physical Society},
  doi = {10.1103/PhysRevB.98.205136},
  url = {https://link.aps.org/doi/10.1103/PhysRevB.98.205136}
}

@article{PhysRevB.108.054307,
  title = {Critical phase and spin sharpening in SU(2)-symmetric monitored quantum circuits},
  author = {Majidy, Shayan and Agrawal, Utkarsh and Gopalakrishnan, Sarang and Potter, Andrew C. and Vasseur, Romain and Halpern, Nicole Yunger},
  journal = {Phys. Rev. B},
  volume = {108},
  issue = {5},
  pages = {054307},
  numpages = {13},
  year = {2023},
  month = {Aug},
  publisher = {American Physical Society},
  doi = {10.1103/PhysRevB.108.054307},
  url = {https://link.aps.org/doi/10.1103/PhysRevB.108.054307}
}

@article{PhysRevLett.131.210402,
  title = {Full Counting Statistics of Charge in Chaotic Many-Body Quantum Systems},
  author = {McCulloch, Ewan and De Nardis, Jacopo and Gopalakrishnan, Sarang and Vasseur, Romain},
  journal = {Phys. Rev. Lett.},
  volume = {131},
  issue = {21},
  pages = {210402},
  numpages = {7},
  year = {2023},
  month = {Nov},
  publisher = {American Physical Society},
  doi = {10.1103/PhysRevLett.131.210402},
  url = {https://link.aps.org/doi/10.1103/PhysRevLett.131.210402}
}

@article{PhysRevLett.56.746,
  title = {Universal term in the free energy at a critical point and the conformal anomaly},
  author = {Affleck, Ian},
  journal = {Phys. Rev. Lett.},
  volume = {56},
  issue = {7},
  pages = {746--748},
  numpages = {0},
  year = {1986},
  month = {Feb},
  publisher = {American Physical Society},
  doi = {10.1103/PhysRevLett.56.746},
  url = {https://link.aps.org/doi/10.1103/PhysRevLett.56.746}
}

@article{PhysRevLett.56.742,
  title = {Conformal invariance, the central charge, and universal finite-size amplitudes at criticality},
  author = {Bl\"ote, H. W. J. and Cardy, John L. and Nightingale, M. P.},
  journal = {Phys. Rev. Lett.},
  volume = {56},
  issue = {7},
  pages = {742--745},
  numpages = {0},
  year = {1986},
  month = {Feb},
  publisher = {American Physical Society},
  doi = {10.1103/PhysRevLett.56.742},
  url = {https://link.aps.org/doi/10.1103/PhysRevLett.56.742}
}

@article{PhysRevB.110.045135,
  title = {Charge and entanglement criticality in a U(1)-symmetric hybrid circuit of qubits},
  author = {Chakraborty, Ahana and Chen, Kun and Zabalo, Aidan and Wilson, Justin H. and Pixley, J. H.},
  journal = {Phys. Rev. B},
  volume = {110},
  issue = {4},
  pages = {045135},
  numpages = {17},
  year = {2024},
  month = {Jul},
  publisher = {American Physical Society},
  doi = {10.1103/PhysRevB.110.045135},
  url = {https://link.aps.org/doi/10.1103/PhysRevB.110.045135}
}

@article{MSR1,
  title = {Statistical Dynamics of Classical Systems},
  author = {Martin, P. C. and Siggia, E. D. and Rose, H. A.},
  journal = {Phys. Rev. A},
  volume = {8},
  issue = {1},
  pages = {423--437},
  numpages = {0},
  year = {1973},
  month = {Jul},
  publisher = {American Physical Society},
  doi = {10.1103/PhysRevA.8.423},
  url = {https://link.aps.org/doi/10.1103/PhysRevA.8.423}
}

@article{MSR2,
  title = {Field-theory renormalization and critical dynamics above ${T}_{c}$: Helium, antiferromagnets, and liquid-gas systems},
  author = {De Dominicis, C. and Peliti, L.},
  journal = {Phys. Rev. B},
  volume = {18},
  issue = {1},
  pages = {353--376},
  numpages = {0},
  year = {1978},
  month = {Jul},
  publisher = {American Physical Society},
  doi = {10.1103/PhysRevB.18.353},
  url = {https://link.aps.org/doi/10.1103/PhysRevB.18.353}
}

@article{MSR3,
	author = {Janssen, Hans-Karl},
	date = {1976/12/01},
	doi = {10.1007/BF01316547},
	id = {Janssen1976},
	isbn = {1431-584X},
	journal = {Zeitschrift f{\"u}r Physik B Condensed Matter},
	number = {4},
	pages = {377--380},
	title = {On a Lagrangean for classical field dynamics and renormalization group calculations of dynamical critical properties},
	url = {https://doi.org/10.1007/BF01316547},
	volume = {23},
	year = {1976}}

@article{PhysRevB.103.174309,
  title = {Self-organized error correction in random unitary circuits with measurement},
  author = {Fan, Ruihua and Vijay, Sagar and Vishwanath, Ashvin and You, Yi-Zhuang},
  journal = {Phys. Rev. B},
  volume = {103},
  issue = {17},
  pages = {174309},
  numpages = {26},
  year = {2021},
  month = {May},
  publisher = {American Physical Society},
  doi = {10.1103/PhysRevB.103.174309},
  url = {https://link.aps.org/doi/10.1103/PhysRevB.103.174309}
}

@article{PhysRevB.99.224307,
  title = {Unitary-projective entanglement dynamics},
  author = {Chan, Amos and Nandkishore, Rahul M. and Pretko, Michael and Smith, Graeme},
  journal = {Phys. Rev. B},
  volume = {99},
  issue = {22},
  pages = {224307},
  numpages = {16},
  year = {2019},
  month = {Jun},
  publisher = {American Physical Society},
  doi = {10.1103/PhysRevB.99.224307},
  url = {https://link.aps.org/doi/10.1103/PhysRevB.99.224307}
}

@article{PhysRevB.104.104305,
  title = {Conformal invariance and quantum nonlocality in critical hybrid circuits},
  author = {Li, Yaodong and Chen, Xiao and Ludwig, Andreas W. W. and Fisher, Matthew P. A.},
  journal = {Phys. Rev. B},
  volume = {104},
  issue = {10},
  pages = {104305},
  numpages = {28},
  year = {2021},
  month = {Sep},
  publisher = {American Physical Society},
  doi = {10.1103/PhysRevB.104.104305},
  url = {https://link.aps.org/doi/10.1103/PhysRevB.104.104305}
}

@article{PhysRevB.100.064204,
  title = {Entanglement transition from variable-strength weak measurements},
  author = {Szyniszewski, M. and Romito, A. and Schomerus, H.},
  journal = {Phys. Rev. B},
  volume = {100},
  issue = {6},
  pages = {064204},
  numpages = {8},
  year = {2019},
  month = {Aug},
  publisher = {American Physical Society},
  doi = {10.1103/PhysRevB.100.064204},
  url = {https://link.aps.org/doi/10.1103/PhysRevB.100.064204}
}

@article{PhysRevX.10.011042,
  title = {Subdiffusion and Heat Transport in a Tilted Two-Dimensional Fermi-Hubbard System},
  author = {Guardado-Sanchez, Elmer and Morningstar, Alan and Spar, Benjamin M. and Brown, Peter T. and Huse, David A. and Bakr, Waseem S.},
  journal = {Phys. Rev. X},
  volume = {10},
  issue = {1},
  pages = {011042},
  numpages = {8},
  year = {2020},
  month = {Feb},
  publisher = {American Physical Society},
  doi = {10.1103/PhysRevX.10.011042},
  url = {https://link.aps.org/doi/10.1103/PhysRevX.10.011042}
}

@inbook{Levy,
author = {Chechkin, Alexei V. and Metzler, Ralf and Klafter, Joseph and Gonchar, Vsevolod Yu.},
publisher = {John Wiley and Sons, Ltd},
isbn = {9783527622979},
title = {Introduction to the Theory of Lévy Flights},
booktitle = {Anomalous Transport},
chapter = {5},
pages = {129-162},
doi = {https://doi.org/10.1002/9783527622979.ch5},
year = {2008}
}

@misc{nahum2025bayesiancriticalpointsclassical,
      title={Bayesian critical points in classical lattice models}, 
      author={Adam Nahum and Jesper Lykke Jacobsen},
      year={2025},
      eprint={2504.01264},
      archivePrefix={arXiv},
      primaryClass={cond-mat.stat-mech},
      url={https://arxiv.org/abs/2504.01264}, 
}

@ARTICLE{2025arXiv250412385P,
       author = {{P{\"u}tz}, Malte and {Garratt}, Samuel J. and {Nishimori}, Hidetoshi and {Trebst}, Simon and {Zhu}, Guo-Yi},
        title = "{Learning transitions in classical Ising models and deformed toric codes}",
      journal = {arXiv e-prints},
         year = 2025,
        month = apr,
          eid = {arXiv:2504.12385},
        pages = {arXiv:2504.12385},
          doi = {10.48550/arXiv.2504.12385},
archivePrefix = {arXiv},
       eprint = {2504.12385},
 primaryClass = {cond-mat.stat-mech},
       adsurl = {https://ui.adsabs.harvard.edu/abs/2025arXiv250412385P}
}

@ARTICLE{2025arXiv250408888K,
       author = {{Kim}, Sun Woo P. and {von Keyserlingk}, Curt and {Lamacraft}, Austen},
        title = "{Measurement-induced phase transitions in quantum inference problems and quantum hidden Markov models}",
      journal = {arXiv e-prints},
         year = 2025,
        month = apr,
          eid = {arXiv:2504.08888},
        pages = {arXiv:2504.08888},
          doi = {10.48550/arXiv.2504.08888},
archivePrefix = {arXiv},
       eprint = {2504.08888},
 primaryClass = {cond-mat.stat-mech},
       adsurl = {https://ui.adsabs.harvard.edu/abs/2025arXiv250408888K}
}

@ARTICLE{2024PhRvB.110o5150S,
       author = {{Sala}, Pablo and {Gopalakrishnan}, Sarang and {Oshikawa}, Masaki and {You}, Yizhi},
        title = "{Spontaneous strong symmetry breaking in open systems: Purification perspective}",
      journal = {\prb},
         year = 2024,
        month = oct,
       volume = {110},
       number = {15},
          eid = {155150},
        pages = {155150},
          doi = {10.1103/PhysRevB.110.155150},
archivePrefix = {arXiv},
       eprint = {2405.02402},
 primaryClass = {quant-ph},
       adsurl = {https://ui.adsabs.harvard.edu/abs/2024PhRvB.110o5150S}
}

@article{PRXQuantum.6.010344,
  title = {Strong-to-Weak Spontaneous Symmetry Breaking in Mixed Quantum States},
  author = {Lessa, Leonardo A. and Ma, Ruochen and Zhang, Jian-Hao and Bi, Zhen and Cheng, Meng and Wang, Chong},
  journal = {PRX Quantum},
  volume = {6},
  issue = {1},
  pages = {010344},
  numpages = {24},
  year = {2025},
  month = {Mar},
  publisher = {American Physical Society},
  doi = {10.1103/PRXQuantum.6.010344},
  url = {https://link.aps.org/doi/10.1103/PRXQuantum.6.010344}
}

@ARTICLE{2023arXiv230516399M,
       author = {{Ma}, Ruochen and {Zhang}, Jian-Hao and {Bi}, Zhen and {Cheng}, Meng and {Wang}, Chong},
        title = "{Topological Phases with Average Symmetries: the Decohered, the Disordered, and the Intrinsic}",
      journal = {arXiv e-prints},
         year = 2023,
        month = may,
          eid = {arXiv:2305.16399},
        pages = {arXiv:2305.16399},
          doi = {10.48550/arXiv.2305.16399},
archivePrefix = {arXiv},
       eprint = {2305.16399},
 primaryClass = {cond-mat.str-el},
       adsurl = {https://ui.adsabs.harvard.edu/abs/2023arXiv230516399M}
}

@article{PRXQuantum.4.030317,
  title = {Quantum Criticality Under Decoherence or Weak Measurement},
  author = {Lee, Jong Yeon and Jian, Chao-Ming and Xu, Cenke},
  journal = {PRX Quantum},
  volume = {4},
  issue = {3},
  pages = {030317},
  numpages = {20},
  year = {2023},
  month = {Aug},
  publisher = {American Physical Society},
  doi = {10.1103/PRXQuantum.4.030317},
  url = {https://link.aps.org/doi/10.1103/PRXQuantum.4.030317}
}

@article{PhysRevResearch.2.033124,
  title = {Fracton hydrodynamics},
  author = {Gromov, Andrey and Lucas, Andrew and Nandkishore, Rahul M.},
  journal = {Phys. Rev. Res.},
  volume = {2},
  issue = {3},
  pages = {033124},
  numpages = {11},
  year = {2020},
  month = {Jul},
  publisher = {American Physical Society},
  doi = {10.1103/PhysRevResearch.2.033124},
  url = {https://link.aps.org/doi/10.1103/PhysRevResearch.2.033124}
}

@article{PhysRevLett.125.245303,
  title = {Anomalous Diffusion in Dipole- and Higher-Moment-Conserving Systems},
  author = {Feldmeier, Johannes and Sala, Pablo and De Tomasi, Giuseppe and Pollmann, Frank and Knap, Michael},
  journal = {Phys. Rev. Lett.},
  volume = {125},
  issue = {24},
  pages = {245303},
  numpages = {6},
  year = {2020},
  month = {Dec},
  publisher = {American Physical Society},
  doi = {10.1103/PhysRevLett.125.245303},
  url = {https://link.aps.org/doi/10.1103/PhysRevLett.125.245303}
}

@article{agrawal2023observingquantummeasurementcollapse,
  title = {Observing Quantum Measurement Collapse as a Learnability Phase Transition},
  author = {Agrawal, Utkarsh and Lopez-Piqueres, Javier and Vasseur, Romain and Gopalakrishnan, Sarang and Potter, Andrew C.},
  journal = {Phys. Rev. X},
  volume = {14},
  issue = {4},
  pages = {041012},
  numpages = {12},
  year = {2024},
  month = {Oct},
  publisher = {American Physical Society},
  doi = {10.1103/PhysRevX.14.041012},
  url = {https://link.aps.org/doi/10.1103/PhysRevX.14.041012}
}

@article{Li_2023,
   title={Entanglement Domain Walls in Monitored Quantum Circuits and the Directed Polymer in a Random Environment},
   volume={4},
   ISSN={2691-3399},
   url={http://dx.doi.org/10.1103/PRXQuantum.4.010331},
   DOI={10.1103/prxquantum.4.010331},
   number={1},
   journal={PRX Quantum},
   publisher={American Physical Society (APS)},
   author={Li, Yaodong and Vijay, Sagar and Fisher, Matthew P.A.},
   year={2023},
   month=mar }

@inbook{Potter_2022,
   title={Entanglement Dynamics in Hybrid Quantum Circuits},
   ISBN={9783031039980},
   ISSN={2364-9062},
   url={http://dx.doi.org/10.1007/978-3-031-03998-0_9},
   DOI={10.1007/978-3-031-03998-0_9},
   booktitle={Entanglement in Spin Chains},
   publisher={Springer International Publishing},
   author={Potter, Andrew C. and Vasseur, Romain},
   year={2022},
   pages={211–249} }

@article{PhysRevX.11.011030,
  title = {Entanglement Phase Transitions in Measurement-Only Dynamics},
  author = {Ippoliti, Matteo and Gullans, Michael J. and Gopalakrishnan, Sarang and Huse, David A. and Khemani, Vedika},
  journal = {Phys. Rev. X},
  volume = {11},
  issue = {1},
  pages = {011030},
  numpages = {23},
  year = {2021},
  month = {Feb},
  publisher = {American Physical Society},
  doi = {10.1103/PhysRevX.11.011030},
  url = {https://link.aps.org/doi/10.1103/PhysRevX.11.011030}
}

@article{lavasani_measurement-induced_2021,
	title = {Measurement-induced topological entanglement transitions in symmetric random quantum circuits},
	volume = {17},
	issn = {1745-2481},
	url = {https://doi.org/10.1038/s41567-020-01112-z},
	doi = {10.1038/s41567-020-01112-z},
	number = {3},
	journal = {Nature Physics},
	author = {Lavasani, Ali and Alavirad, Yahya and Barkeshli, Maissam},
	month = mar,
	year = {2021},
	pages = {342--347},
}

@article{PhysRevResearch.3.023200,
  title = {Measurement-protected quantum phases},
  author = {Sang, Shengqi and Hsieh, Timothy H.},
  journal = {Phys. Rev. Res.},
  volume = {3},
  issue = {2},
  pages = {023200},
  numpages = {9},
  year = {2021},
  month = {Jun},
  publisher = {American Physical Society},
  doi = {10.1103/PhysRevResearch.3.023200},
  url = {https://link.aps.org/doi/10.1103/PhysRevResearch.3.023200}
}

@article{PhysRevB.103.224210,
  title = {Measurement-induced entanglement transitions in the quantum Ising chain: From infinite to zero clicks},
  author = {Turkeshi, Xhek and Biella, Alberto and Fazio, Rosario and Dalmonte, Marcello and Schir\'o, Marco},
  journal = {Phys. Rev. B},
  volume = {103},
  issue = {22},
  pages = {224210},
  numpages = {13},
  year = {2021},
  month = {Jun},
  publisher = {American Physical Society},
  doi = {10.1103/PhysRevB.103.224210},
  url = {https://link.aps.org/doi/10.1103/PhysRevB.103.224210}
}

@article{PhysRevLett.126.060501,
  title = {Postselection-Free Entanglement Dynamics via Spacetime Duality},
  author = {Ippoliti, Matteo and Khemani, Vedika},
  journal = {Phys. Rev. Lett.},
  volume = {126},
  issue = {6},
  pages = {060501},
  numpages = {7},
  year = {2021},
  month = {Feb},
  publisher = {American Physical Society},
  doi = {10.1103/PhysRevLett.126.060501},
  url = {https://link.aps.org/doi/10.1103/PhysRevLett.126.060501}
}

@article{PRXQuantum.2.040319,
  title = {Spacetime duality between localization transitions and measurement-induced transitions},
  author = {Lu, Tsung-Cheng and Grover, Tarun},
  journal = {PRX Quantum},
  volume = {2},
  issue = {4},
  pages = {040319},
  numpages = {17},
  year = {2021},
  month = {Oct},
  publisher = {American Physical Society},
  doi = {10.1103/PRXQuantum.2.040319},
  url = {https://link.aps.org/doi/10.1103/PRXQuantum.2.040319}
}

@article{PhysRevB.106.134206,
  title = {Criticality and entanglement in nonunitary quantum circuits and tensor networks of noninteracting fermions},
  author = {Jian, Chao-Ming and Bauer, Bela and Keselman, Anna and Ludwig, Andreas W. W.},
  journal = {Phys. Rev. B},
  volume = {106},
  issue = {13},
  pages = {134206},
  numpages = {38},
  year = {2022},
  month = {Oct},
  publisher = {American Physical Society},
  doi = {10.1103/PhysRevB.106.134206},
  url = {https://link.aps.org/doi/10.1103/PhysRevB.106.134206}
}

@article{PhysRevLett.126.170503,
  title = {Entanglement and Purification Transitions in Non-Hermitian Quantum Mechanics},
  author = {Gopalakrishnan, Sarang and Gullans, Michael J.},
  journal = {Phys. Rev. Lett.},
  volume = {126},
  issue = {17},
  pages = {170503},
  numpages = {6},
  year = {2021},
  month = {Apr},
  publisher = {American Physical Society},
  doi = {10.1103/PhysRevLett.126.170503},
  url = {https://link.aps.org/doi/10.1103/PhysRevLett.126.170503}
}

@article{BAO2021168618,
title = {Symmetry enriched phases of quantum circuits},
journal = {Annals of Physics},
volume = {435},
pages = {168618},
year = {2021},
note = {Special issue on Philip W. Anderson},
issn = {0003-4916},
doi = {https://doi.org/10.1016/j.aop.2021.168618},
url = {https://www.sciencedirect.com/science/article/pii/S0003491621002244},
author = {Yimu Bao and Soonwon Choi and Ehud Altman}
}

@article{PhysRevLett.128.010604,
  title = {Measurement-Induced Transition in Long-Range Interacting Quantum Circuits},
  author = {Block, Maxwell and Bao, Yimu and Choi, Soonwon and Altman, Ehud and Yao, Norman Y.},
  journal = {Phys. Rev. Lett.},
  volume = {128},
  issue = {1},
  pages = {010604},
  numpages = {7},
  year = {2022},
  month = {Jan},
  publisher = {American Physical Society},
  doi = {10.1103/PhysRevLett.128.010604},
  url = {https://link.aps.org/doi/10.1103/PhysRevLett.128.010604}
}

@article{PhysRevB.104.094304,
  title = {Measurement-induced purification in large-$N$ hybrid Brownian circuits},
  author = {Bentsen, Gregory S. and Sahu, Subhayan and Swingle, Brian},
  journal = {Phys. Rev. B},
  volume = {104},
  issue = {9},
  pages = {094304},
  numpages = {32},
  year = {2021},
  month = {Sep},
  publisher = {American Physical Society},
  doi = {10.1103/PhysRevB.104.094304},
  url = {https://link.aps.org/doi/10.1103/PhysRevB.104.094304}
}

@article{PhysRevResearch.4.023146,
  title = {Generalized quantum measurements with matrix product states: Entanglement phase transition and clusterization},
  author = {Doggen, Elmer V. H. and Gefen, Yuval and Gornyi, Igor V. and Mirlin, Alexander D. and Polyakov, Dmitry G.},
  journal = {Phys. Rev. Res.},
  volume = {4},
  issue = {2},
  pages = {023146},
  numpages = {24},
  year = {2022},
  month = {May},
  publisher = {American Physical Society},
  doi = {10.1103/PhysRevResearch.4.023146},
  url = {https://link.aps.org/doi/10.1103/PhysRevResearch.4.023146}
}

@article{PhysRevLett.128.130605,
  title = {Universal Behavior beyond Multifractality of Wave Functions at Measurement-Induced Phase Transitions},
  author = {Sierant, Piotr and Turkeshi, Xhek},
  journal = {Phys. Rev. Lett.},
  volume = {128},
  issue = {13},
  pages = {130605},
  numpages = {7},
  year = {2022},
  month = {Apr},
  publisher = {American Physical Society},
  doi = {10.1103/PhysRevLett.128.130605},
  url = {https://link.aps.org/doi/10.1103/PhysRevLett.128.130605}
}

@article{RevModPhys.87.593,
  title = {Macroscopic fluctuation theory},
  author = {Bertini, Lorenzo and De Sole, Alberto and Gabrielli, Davide and Jona-Lasinio, Giovanni and Landim, Claudio},
  journal = {Rev. Mod. Phys.},
  volume = {87},
  issue = {2},
  pages = {593--636},
  numpages = {44},
  year = {2015},
  month = {Jun},
  publisher = {American Physical Society},
  doi = {10.1103/RevModPhys.87.593},
  url = {https://link.aps.org/doi/10.1103/RevModPhys.87.593}
}

@article{PhysRevLett.130.220404,
  title = {Cross Entropy Benchmark for Measurement-Induced Phase Transitions},
  author = {Li, Yaodong and Zou, Yijian and Glorioso, Paolo and Altman, Ehud and Fisher, Matthew P. A.},
  journal = {Phys. Rev. Lett.},
  volume = {130},
  issue = {22},
  pages = {220404},
  numpages = {6},
  year = {2023},
  month = {Jun},
  publisher = {American Physical Society},
  doi = {10.1103/PhysRevLett.130.220404},
  url = {https://link.aps.org/doi/10.1103/PhysRevLett.130.220404}
}

@article{PhysRevB.109.125148,
  title = {Measurement-induced phase transitions in the toric code},
  author = {Negari, Amir-Reza and Sahu, Subhayan and Hsieh, Timothy H.},
  journal = {Phys. Rev. B},
  volume = {109},
  issue = {12},
  pages = {125148},
  numpages = {13},
  year = {2024},
  month = {Mar},
  publisher = {American Physical Society},
  doi = {10.1103/PhysRevB.109.125148},
  url = {https://link.aps.org/doi/10.1103/PhysRevB.109.125148}
}

@article{PRXQuantum.4.030333,
  title = {Measurement-Induced Phase Transitions on Dynamical Quantum Trees},
  author = {Feng, Xiaozhou and Skinner, Brian and Nahum, Adam},
  journal = {PRX Quantum},
  volume = {4},
  issue = {3},
  pages = {030333},
  numpages = {29},
  year = {2023},
  month = {Sep},
  publisher = {American Physical Society},
  doi = {10.1103/PRXQuantum.4.030333},
  url = {https://link.aps.org/doi/10.1103/PRXQuantum.4.030333}
}

@article{PhysRevB.110.L060202,
  title = {Measurement-induced phase transition in a single-body tight-binding model},
  author = {Jin, Tony and Martin, David G.},
  journal = {Phys. Rev. B},
  volume = {110},
  issue = {6},
  pages = {L060202},
  numpages = {6},
  year = {2024},
  month = {Aug},
  publisher = {American Physical Society},
  doi = {10.1103/PhysRevB.110.L060202},
  url = {https://link.aps.org/doi/10.1103/PhysRevB.110.L060202}
}

@article{PhysRevB.107.L220204,
  title = {Infinite-randomness criticality in monitored quantum dynamics with static disorder},
  author = {Zabalo, Aidan and Wilson, Justin H. and Gullans, Michael J. and Vasseur, Romain and Gopalakrishnan, Sarang and Huse, David A. and Pixley, J. H.},
  journal = {Phys. Rev. B},
  volume = {107},
  issue = {22},
  pages = {L220204},
  numpages = {7},
  year = {2023},
  month = {Jun},
  publisher = {American Physical Society},
  doi = {10.1103/PhysRevB.107.L220204},
  url = {https://link.aps.org/doi/10.1103/PhysRevB.107.L220204}
}

@article{PRXQuantum.5.020304,
  title = {Learnability Transitions in Monitored Quantum Dynamics via Eavesdropper's Classical Shadows},
  author = {Ippoliti, Matteo and Khemani, Vedika},
  journal = {PRX Quantum},
  volume = {5},
  issue = {2},
  pages = {020304},
  numpages = {24},
  year = {2024},
  month = {Apr},
  publisher = {American Physical Society},
  doi = {10.1103/PRXQuantum.5.020304},
  url = {https://link.aps.org/doi/10.1103/PRXQuantum.5.020304}
}

@article{PhysRevB.110.054308,
  title = {Local and nonlocal stochastic control of quantum chaos: Measurement- and control-induced criticality},
  author = {Pan, Haining and Ganeshan, Sriram and Iadecola, Thomas and Wilson, Justin H. and Pixley, J. H.},
  journal = {Phys. Rev. B},
  volume = {110},
  issue = {5},
  pages = {054308},
  numpages = {14},
  year = {2024},
  month = {Aug},
  publisher = {American Physical Society},
  doi = {10.1103/PhysRevB.110.054308},
  url = {https://link.aps.org/doi/10.1103/PhysRevB.110.054308}
}

@article{PhysRevLett.130.120402,
  title = {Controlling Entanglement at Absorbing State Phase Transitions in Random Circuits},
  author = {Sierant, Piotr and Turkeshi, Xhek},
  journal = {Phys. Rev. Lett.},
  volume = {130},
  issue = {12},
  pages = {120402},
  numpages = {8},
  year = {2023},
  month = {Mar},
  publisher = {American Physical Society},
  doi = {10.1103/PhysRevLett.130.120402},
  url = {https://link.aps.org/doi/10.1103/PhysRevLett.130.120402}
}

@article{PhysRevLett.132.240402,
  title = {Entanglement Structure and Information Protection in Noisy Hybrid Quantum Circuits},
  author = {Liu, Shuo and Li, Ming-Rui and Zhang, Shi-Xin and Jian, Shao-Kai},
  journal = {Phys. Rev. Lett.},
  volume = {132},
  issue = {24},
  pages = {240402},
  numpages = {8},
  year = {2024},
  month = {Jun},
  publisher = {American Physical Society},
  doi = {10.1103/PhysRevLett.132.240402},
  url = {https://link.aps.org/doi/10.1103/PhysRevLett.132.240402}
}

@article{PhysRevB.108.L041103,
  title = {Entanglement steering in adaptive circuits with feedback},
  author = {Ravindranath, Vikram and Han, Yiqiu and Yang, Zhi-Cheng and Chen, Xiao},
  journal = {Phys. Rev. B},
  volume = {108},
  issue = {4},
  pages = {L041103},
  numpages = {5},
  year = {2023},
  month = {Jul},
  publisher = {American Physical Society},
  doi = {10.1103/PhysRevB.108.L041103},
  url = {https://link.aps.org/doi/10.1103/PhysRevB.108.L041103}
}

@article{PhysRevB.110.064323,
  title = {Noise-induced phase transitions in hybrid quantum circuits},
  author = {Liu, Shuo and Li, Ming-Rui and Zhang, Shi-Xin and Jian, Shao-Kai and Yao, Hong},
  journal = {Phys. Rev. B},
  volume = {110},
  issue = {6},
  pages = {064323},
  numpages = {13},
  year = {2024},
  month = {Aug},
  publisher = {American Physical Society},
  doi = {10.1103/PhysRevB.110.064323},
  url = {https://link.aps.org/doi/10.1103/PhysRevB.110.064323}
}

@article{PhysRevB.107.014308,
  title = {Charge fluctuation and charge-resolved entanglement in a monitored quantum circuit with $U(1)$ symmetry},
  author = {Oshima, Hisanori and Fuji, Yohei},
  journal = {Phys. Rev. B},
  volume = {107},
  issue = {1},
  pages = {014308},
  numpages = {20},
  year = {2023},
  month = {Jan},
  publisher = {American Physical Society},
  doi = {10.1103/PhysRevB.107.014308},
  url = {https://link.aps.org/doi/10.1103/PhysRevB.107.014308}
}

@article{PhysRevB.107.224303,
  title = {Triviality of quantum trajectories close to a directed percolation transition},
  author = {Piroli, Lorenzo and Li, Yaodong and Vasseur, Romain and Nahum, Adam},
  journal = {Phys. Rev. B},
  volume = {107},
  issue = {22},
  pages = {224303},
  numpages = {16},
  year = {2023},
  month = {Jun},
  publisher = {American Physical Society},
  doi = {10.1103/PhysRevB.107.224303},
  url = {https://link.aps.org/doi/10.1103/PhysRevB.107.224303}
}

@article{PhysRevLett.131.060403,
  title = {Measurement and Feedback Driven Entanglement Transition in the Probabilistic Control of Chaos},
  author = {Iadecola, Thomas and Ganeshan, Sriram and Pixley, J. H. and Wilson, Justin H.},
  journal = {Phys. Rev. Lett.},
  volume = {131},
  issue = {6},
  pages = {060403},
  numpages = {7},
  year = {2023},
  month = {Aug},
  publisher = {American Physical Society},
  doi = {10.1103/PhysRevLett.131.060403},
  url = {https://link.aps.org/doi/10.1103/PhysRevLett.131.060403}
}

@article{PhysRevLett.131.220404,
  title = {Scrambling Transition in a Radiative Random Unitary Circuit},
  author = {Weinstein, Zack and Kelly, Shane P. and Marino, Jamir and Altman, Ehud},
  journal = {Phys. Rev. Lett.},
  volume = {131},
  issue = {22},
  pages = {220404},
  numpages = {7},
  year = {2023},
  month = {Nov},
  publisher = {American Physical Society},
  doi = {10.1103/PhysRevLett.131.220404},
  url = {https://link.aps.org/doi/10.1103/PhysRevLett.131.220404}
}

@article{PhysRevB.108.104203,
  title = {Renormalization group for measurement and entanglement phase transitions},
  author = {Nahum, Adam and Wiese, Kay J\"org},
  journal = {Phys. Rev. B},
  volume = {108},
  issue = {10},
  pages = {104203},
  numpages = {28},
  year = {2023},
  month = {Sep},
  publisher = {American Physical Society},
  doi = {10.1103/PhysRevB.108.104203},
  url = {https://link.aps.org/doi/10.1103/PhysRevB.108.104203}
}

@article{PhysRevB.107.064303,
  title = {Entanglement transitions with free fermions},
  author = {Merritt, Joseph and Fidkowski, Lukasz},
  journal = {Phys. Rev. B},
  volume = {107},
  issue = {6},
  pages = {064303},
  numpages = {9},
  year = {2023},
  month = {Feb},
  publisher = {American Physical Society},
  doi = {10.1103/PhysRevB.107.064303},
  url = {https://link.aps.org/doi/10.1103/PhysRevB.107.064303}
}

@article{PhysRevB.108.L020306,
  title = {Purification timescales in monitored fermions},
  author = {L\'oio, Hugo and De Luca, Andrea and De Nardis, Jacopo and Turkeshi, Xhek},
  journal = {Phys. Rev. B},
  volume = {108},
  issue = {2},
  pages = {L020306},
  numpages = {7},
  year = {2023},
  month = {Jul},
  publisher = {American Physical Society},
  doi = {10.1103/PhysRevB.108.L020306},
  url = {https://link.aps.org/doi/10.1103/PhysRevB.108.L020306}
}

@misc{JianShapourianBauerLudwig2023,
      title={Measurement-induced entanglement transitions in quantum circuits of non-interacting fermions: Born-rule versus forced measurements}, 
      author={Chao-Ming Jian and Hassan Shapourian and Bela Bauer and Andreas W. W. Ludwig},
      year={2023},
      eprint={2302.09094},
      archivePrefix={arXiv},
      primaryClass={cond-mat.stat-mech},
      url={https://arxiv.org/abs/2302.09094}, 
}

@article{PhysRevX.13.041045,
  title = {Nonlinear Sigma Models for Monitored Dynamics of Free Fermions},
  author = {Fava, Michele and Piroli, Lorenzo and Swann, Tobias and Bernard, Denis and Nahum, Adam},
  journal = {Phys. Rev. X},
  volume = {13},
  issue = {4},
  pages = {041045},
  numpages = {33},
  year = {2023},
  month = {Dec},
  publisher = {American Physical Society},
  doi = {10.1103/PhysRevX.13.041045},
  url = {https://link.aps.org/doi/10.1103/PhysRevX.13.041045}
}

@misc{GuoJianFosterLudwigKeldysh2024,
      title={Field theory of monitored, interacting fermion dynamics with charge conservation}, 
      author={Haoyu Guo and Matthew S. Foster and Chao-Ming Jian and Andreas W. W. Ludwig},
      year={2024},
      eprint={2410.07317},
      archivePrefix={arXiv},
      primaryClass={cond-mat.stat-mech},
      url={https://arxiv.org/abs/2410.07317}, 
}

@book{GiamarchiBook,
    author = {Giamarchi, Thierry},
    title = {Quantum Physics in One Dimension},
    publisher = {Oxford University Press},
    year = {2003},
    month = {12},
    isbn = {9780198525004},
    doi = {10.1093/acprof:oso/9780198525004.001.0001},
    url = {https://doi.org/10.1093/acprof:oso/9780198525004.001.0001},
}

@misc{MirlinGornyiEtAlKeldysh2024,
      title={Measurement-induced transitions for interacting fermions}, 
      author={Igor Poboiko and Paul Pöpperl and Igor V. Gornyi and Alexander D. Mirlin},
      year={2024},
      eprint={2410.07334},
      archivePrefix={arXiv},
      primaryClass={quant-ph},
      url={https://arxiv.org/abs/2410.07334}, 
}

@ARTICLE{2025arXiv250100547G,
       author = {{Gerbino}, Federico and {Giachetti}, Guido and {Le Doussal}, Pierre and {De Luca}, Andrea},
        title = "{Measurement-Induced Phase Transition in State Estimation of Chaotic Systems and the Directed Polymer}",
      journal = {arXiv e-prints},
         year = 2024,
        month = dec,
          eid = {arXiv:2501.00547},
        pages = {arXiv:2501.00547},
          doi = {10.48550/arXiv.2501.00547},
archivePrefix = {arXiv},
       eprint = {2501.00547},
 primaryClass = {cond-mat.stat-mech},
       adsurl = {https://ui.adsabs.harvard.edu/abs/2025arXiv250100547G}
}

@ARTICLE{2022PhRvB.106b4305W,
       author = {{Willsher}, Josef and {Liu}, Shu-Wei and {Moessner}, Roderich and {Knolle}, Johannes},
        title = "{Measurement-induced phase transition in a chaotic classical many-body system}",
      journal = {\prb},
         year = 2022,
        month = jul,
       volume = {106},
       number = {2},
          eid = {024305},
        pages = {024305},
          doi = {10.1103/PhysRevB.106.024305},
archivePrefix = {arXiv},
       eprint = {2203.11303},
 primaryClass = {cond-mat.stat-mech},
       adsurl = {https://ui.adsabs.harvard.edu/abs/2022PhRvB.106b4305W}
}

@ARTICLE{2022PhRvB.106u4303P,
       author = {{Pizzi}, Andrea and {Malz}, Daniel and {Nunnenkamp}, Andreas and {Knolle}, Johannes},
        title = "{Bridging the gap between classical and quantum many-body information dynamics}",
      journal = {\prb},
         year = 2022,
        month = dec,
       volume = {106},
       number = {21},
          eid = {214303},
        pages = {214303},
          doi = {10.1103/PhysRevB.106.214303},
archivePrefix = {arXiv},
       eprint = {2204.03016},
 primaryClass = {cond-mat.stat-mech},
       adsurl = {https://ui.adsabs.harvard.edu/abs/2022PhRvB.106u4303P}
}

@ARTICLE{2010PhRvE..82c6702J,
       author = {{Johnson}, T.~H. and {Clark}, S.~R. and {Jaksch}, D.},
        title = "{Dynamical simulations of classical stochastic systems using matrix product states}",
      journal = {\pre},
         year = 2010,
        month = sep,
       volume = {82},
       number = {3},
          eid = {036702},
        pages = {036702},
          doi = {10.1103/PhysRevE.82.036702},
archivePrefix = {arXiv},
       eprint = {1006.2639},
 primaryClass = {cond-mat.stat-mech},
       adsurl = {https://ui.adsabs.harvard.edu/abs/2010PhRvE..82c6702J}
}

@ARTICLE{2009PhR...478....1T,
       author = {{Touchette}, Hugo},
        title = "{The large deviation approach to statistical mechanics}",
      journal = {Phys. Rep.},
         year = 2009,
        month = jul,
       volume = {478},
       number = {1-3},
        pages = {1-69},
          doi = {10.1016/j.physrep.2009.05.002},
archivePrefix = {arXiv},
       eprint = {0804.0327},
 primaryClass = {cond-mat.stat-mech},
       adsurl = {https://ui.adsabs.harvard.edu/abs/2009PhR...478....1T}
}

@ARTICLE{2007PhRvL..98s5702G,
       author = {{Garrahan}, J.~P. and {Jack}, R.~L. and {Lecomte}, V. and {Pitard}, E. and {van Duijvendijk}, K. and {van Wijland}, F.},
        title = "{Dynamical First-Order Phase Transition in Kinetically Constrained Models of Glasses}",
      journal = {\prl},
         year = 2007,
        month = may,
       volume = {98},
       number = {19},
          eid = {195702},
        pages = {195702},
          doi = {10.1103/PhysRevLett.98.195702},
archivePrefix = {arXiv},
       eprint = {cond-mat/0701757},
 primaryClass = {cond-mat.stat-mech},
       adsurl = {https://ui.adsabs.harvard.edu/abs/2007PhRvL..98s5702G}
}

@ARTICLE{2009JPhA...42g5007G,
       author = {{Garrahan}, Juan P. and {Jack}, Robert L. and {Lecomte}, Vivien and {Pitard}, Estelle and {van Duijvendijk}, Kristina and {van Wijland}, Fr{\'e}d{\'e}ric},
        title = "{First-order dynamical phase transition in models of glasses: an approach based on ensembles of histories}",
      journal = {Journal of Physics A Mathematical General},
         year = 2009,
        month = feb,
       volume = {42},
       number = {7},
          eid = {075007},
        pages = {075007},
          doi = {10.1088/1751-8113/42/7/075007},
archivePrefix = {arXiv},
       eprint = {0810.5298},
 primaryClass = {cond-mat.stat-mech},
       adsurl = {https://ui.adsabs.harvard.edu/abs/2009JPhA...42g5007G}
}

@ARTICLE{2007JSP...127...51L,
       author = {{Lecomte}, V. and {Appert-Rolland}, C. and {van Wijland}, F.},
        title = "{Thermodynamic Formalism for Systems with Markov Dynamics}",
      journal = {Journal of Statistical Physics},
         year = 2007,
        month = apr,
       volume = {127},
       number = {1},
        pages = {51-106},
          doi = {10.1007/s10955-006-9254-0},
archivePrefix = {arXiv},
       eprint = {cond-mat/0606211},
 primaryClass = {cond-mat.stat-mech},
       adsurl = {https://ui.adsabs.harvard.edu/abs/2007JSP...127...51L}
}

@ARTICLE{2015PhyA..418...17M,
       author = {{Mallick}, Kirone},
        title = "{The exclusion process: A paradigm for non-equilibrium behaviour}",
      journal = {Physica A Statistical Mechanics and its Applications},
         year = 2015,
        month = jan,
       volume = {418},
        pages = {17-48},
          doi = {10.1016/j.physa.2014.07.046},
archivePrefix = {arXiv},
       eprint = {1412.6258},
 primaryClass = {cond-mat.stat-mech},
       adsurl = {https://ui.adsabs.harvard.edu/abs/2015PhyA..418...17M}
}

@ARTICLE{2012NJPh...14g3007B,
       author = {{Bu{\v{c}}a}, Berislav and {Prosen}, Toma{\v{z}}},
        title = "{A note on symmetry reductions of the Lindblad equation: transport in constrained open spin chains}",
      journal = {New Journal of Physics},
         year = 2012,
        month = jul,
       volume = {14},
       number = {7},
          eid = {073007},
        pages = {073007},
          doi = {10.1088/1367-2630/14/7/073007},
archivePrefix = {arXiv},
       eprint = {1203.0943},
 primaryClass = {quant-ph},
       adsurl = {https://ui.adsabs.harvard.edu/abs/2012NJPh...14g3007B}
}

@article{PhysRevLett.96.120603,
  title = {Direct Evaluation of Large-Deviation Functions},
  author = {Giardin\`a, Cristian and Kurchan, Jorge and Peliti, Luca},
  journal = {Phys. Rev. Lett.},
  volume = {96},
  issue = {12},
  pages = {120603},
  numpages = {4},
  year = {2006},
  month = {Mar},
  publisher = {American Physical Society},
  doi = {10.1103/PhysRevLett.96.120603},
  url = {https://link.aps.org/doi/10.1103/PhysRevLett.96.120603}
}

@ARTICLE{2011JSP...145..787G,
       author = {{Giardina}, Cristian and {Kurchan}, Jorge and {Lecomte}, Vivien and {Tailleur}, Julien},
        title = "{Simulating Rare Events in Dynamical Processes}",
      journal = {Journal of Statistical Physics},
         year = 2011,
        month = nov,
       volume = {145},
       number = {4},
        pages = {787-811},
          doi = {10.1007/s10955-011-0350-4},
archivePrefix = {arXiv},
       eprint = {1106.4929},
 primaryClass = {cond-mat.stat-mech},
       adsurl = {https://ui.adsabs.harvard.edu/abs/2011JSP...145..787G}
}

@dataset{data,
  author       = {McCulloch, Ewan},
  title        = {Simulation data for "Monitored Fluctuation
                   Hydrodynamics"
                  },
  month        = jun,
  year         = 2025,
  publisher    = {Zenodo},
  doi          = {10.5281/zenodo.15620590},
  url          = {https://doi.org/10.5281/zenodo.15620590},
}
\end{document}